\documentclass[pdflatex,sn-nature]{sn-jnl}

\usepackage{graphicx}%
\usepackage{multirow}%
\usepackage{amsmath,amssymb,amsfonts}%
\usepackage{amsthm}%
\usepackage{mathrsfs}%
\usepackage[title]{appendix}%
\usepackage{xcolor}%
\usepackage{textcomp}%
\usepackage{manyfoot}%
\usepackage{booktabs}%
\usepackage{algorithm}%
\usepackage{algorithmicx}%
\usepackage{algpseudocode}%
\usepackage{listings}%
\usepackage{subfigure}
\usepackage{comment}
\usepackage[normalem]{ulem}
 \useunder{\uline}{\ul}{}
 \usepackage{lineno}
 \usepackage{enumitem}
 
\theoremstyle{thmstyleone}%
\theoremstyle{thmstyletwo}%

\theoremstyle{thmstylethree}%

\begin{document}

\title[A Relational Model of Neighborhood Mobility]{A Relational Model of Neighborhood Mobility: The Role of Amenities and Cultural Alignment}

\author*[1]{\fnm{Thiago} \sur{H. Silva}}\email{th.silva@utoronto.ca}
\equalcont{These authors contributed equally to this work.}

\author*[2]{\fnm{Daniel} \sur{Silver}}\email{dan.silver@utoronto.ca}
\equalcont{These authors contributed equally to this work.}

\author[3]{\fnm{Gustavo} \sur{Santos}}\email{gustavohenriquesantos@alunos.utfpr.edu.br}

\author[3]{\fnm{Myriam} \sur{Delgado}}\email{myriamdelg@utfpr.edu.br}

\affil*[1]{\orgdiv{School of Cities}, \orgname{University of Toronto}, \orgaddress{\city{Toronto},  \country{Canada}}}

\affil[2]{\orgdiv{Sociology}, \orgname{University of Toronto}, \orgaddress{\city{Toronto},  \country{Canada}}}

\affil[3]{\orgdiv{Informatics}, \orgname{Federal University of Technology}, \orgaddress{\city{Curitiba}, \country{Brazil}}}


\abstract{Why are some neighborhoods strongly connected while others remain isolated? Although standard explanations focus on demographics, economics, and geography, movement across the city may also depend on cultural styles and amenity mix. This study proposes a relational, cross-national model in which local culture and amenity mix alignment creates a ``soft infrastructure'' of urban mobility, i.e., symbolic cues and functional features that shape expectations about the character of places. Using $\approx$650 million Google Places reviews to measure co-visitation between U.S. ZIP codes and $\approx$30 million Canadian change-of-address to track residential mobility, results show that neighborhoods with similar cultural styles and amenities are significantly more connected. These effects persist even after controlling for race, income, education, politics, housing costs, and distance. Urban cohesion and segregation depend not only on who lives where or how far apart neighborhoods are, but on the shared cultural and material ecologies that structure movement across the city. }


\maketitle

\section*{Introduction }\label{secIntro}

How do complex human societies cohere? The coordination of millions of individuals in dense urban environments represents a fundamental challenge of social organization. While formal institutions, economic markets, and demographic sorting play crucial roles, much of the order and predictability of human activity is also guided by shared cultural meanings and material affordances~\cite{antipova_local_2018}.
Early theorists viewed neighborhoods as interdependent elements of an urban ecology in which cultural, political, social, and economic dimensions overlap \cite{park2019city,abbott2017department}. Contemporary urban sociology similarly emphasizes how demographic and economic ties bind neighborhoods together or separate them, shaping health, crime, and inequality \cite{sampson2012great,gieryn_space_2000}. From this perspective, a neighborhood is constituted not only by its residents and institutions but also by its evolving connections to other places through flows of people, ideas, and investment. 

Standard models explain these connections by focusing on race, income, education, and geographic factors\footnote{In the experiments, all these factors are addressed through similarity metrics. A high value for geographic similarity, for example,  indicates that neighborhoods are located near each other.} \cite{song2010limits,alessandretti2018evidence,candipan2021residence,phillips2021social,saxon2021local}. Related work highlights the role of information, whereby individuals and households make decisions (about where to visit or move, for example)  under uncertain conditions: they often appeal to personal networks, news media, real estate listings, online reviews, and the like \cite{galster2024UrbStd}. Yet far less is known about how the symbolic resonance of a place and the composition of its everyday amenities structure the flows of people that knit the social fabric together. Large-scale empirical analysis of these interdependencies has long been constrained by limited data~\cite{knaap2019dynamics}. 

Recent methodological advances now make it possible to study cities in a more fully relational way using a wider range of variables. Fine-grained, geo-referenced traces from location-based services enable researchers to model neighborhoods through the flows that link them \citep{Phillips2019Social,Silva:2019:UCL:3309872.3301284,Poorthuis2021Changing,Shelton2019Nature}. When combined with network-analytic methods, these data support more precise conceptualizations of urban structure: the extent to which neighborhoods are connected through movements of residents, visitors, or resources. In this framework, cities can be represented as graphs whose vertices are neighborhoods and whose edges capture flows of people, capital, businesses, and ideas \citep{phillips2021social,sampson2012great,papachristos2018connected,Graif2019Network,Candipan2021From,Daepp2021Small,jisaGustavo2025}. This in turn allows researchers to measure not only shared demographic features but also similarities in amenity environments and local culture \cite{wang2018urban,candipan2021residence,SILVER2023104130,de2024people,silvaSilver2024EPB}, as well as how these relationships vary across time. 

A growing body of research demonstrates the promise of this approach. For instance, \cite{Candipan2021From} utilizes geo-tagged Twitter data to construct a segregated mobility index across 50 U.S. cities; \cite{Andris2019Threads} demonstrates how youth mentoring programs facilitate the weaving of neighborhood networks that cross socioeconomic divides; \cite{Shelton2015Social} documents that ostensibly divided Louisville neighborhoods are, in fact, fluidly interconnected; and \cite{jisaGustavo2025} uses Foursquare and Google Places data to show that geographic distance remains the dominant factor structuring mobility between neighborhood pairs. Other work maps how mobility graphs reveal experienced isolation \cite{athey2021estimating}, clusters neighborhoods based on credit-based mobility \cite{Daepp2021Small}, connects disadvantaged areas to more polluted ones \cite{Brazil2022Environmental}, and traces shifting connections during gentrification \cite{Poorthuis2021Changing}. Network-based models have also been incorporated into predictive studies of gentrification, crime, disease dynamics, and local culture \citep{papachristos2018connected,saxon2021local,silvaSilver2024EPB}, underscoring the growing importance of relational approaches to urban space.

Even as modern computational tools have begun to overcome data constraints, progress has been piecemeal: most research has examined only a small set of neighborhood characteristics, focused on a single or a few cities, and relied largely on single time points, yielding static snapshots rather than examining relational patterns across multiple time intervals.  As a result, key gaps persist: many studies analyze only one national context, map connectivity without relating it to other variables, restrict attention to a single type of connection, or characterize neighborhood similarity using a narrow range of economic or demographic indicators. Although understandable, these approaches make it difficult to evaluate how strong a particular relationship is or whether similar patterns hold across national contexts and outcomes. Little work has examined how similarities in local culture and material affordance of local environments shape inter-neighborhood ties, despite these features constituting the everyday environments through which residents navigate urban life. 

In this study, we propose that alignment in amenity mix \cite{gubertAsonam24,jisaGubert2025} (capturing material affordance of local environments)  and cultural style or scenes \cite{silver2016scenescapes,jisaGubert2025} (capturing local culture)  acts as a distinct relational force connecting neighborhoods, complementing the influence of demographic, economic, political, and geographic factors. We conceptualize these two dimensions as a ``soft infrastructure'' of urban mobility: symbolic signals and functional affordances that shape how people perceive, evaluate, and navigate place.

Using two large-scale datasets, co-visitation records derived from  Google Places reviews across U.S. ZIP codes and tax-record-based residential mobility flows across Canadian postal areas, we show that neighborhoods with more similar amenity mixes and cultural styles exhibit significantly stronger mobility ties, even after adjusting for race, income, education, politics, housing costs, and distance. These results suggest that urban cohesion depends not only on population composition or spatial proximity but also on shared cultural and material ecologies that structure inter-neighborhood relations.

Our work contributes to urban and neighborhood research in two main ways. First, we augment existing models of neighborhood connectivity with local cultural scenes, which emphasize the symbolic meaning of places, and amenity mix, highlighting functionality or material affordances of local environments. In doing so, and second, we synthesize three strands of existing research: (1) network-based studies of relations between urban areas \cite{alessandretti2018evidence,kraemer2020mapping,candipan2021residence,de2024people,xu2025using, levy2020triple,phillips2021social,levy2022neighborhood,Huang2023,silvaSilver2024EPB,jisaGustavo2025}; (2) studies of the role of local amenities in orienting flows of people between locations \cite{SILVER2023104130,abbiasov202415,de2024people,xu2025using}; and (3) studies of the heuristics or cues by which individuals or households decide where to visit or move~\cite{bayer2007unified,andersen2017selective,galster2017status,hasan2019digitization,palm2001residential,galster2024UrbStd}. To join and add to this research we a) build a neighborhood relational model that evaluates the role of various factors in driving neighborhood connectivity; b) include culture and amenities as possible cues that signal the type of a place to potential residents and visitors; and span c)  two nations; d) two types of mobility;  e) over multiple time points. The discussion suggests possible mechanisms underlying the findings and proposes avenues for future research to identify them more directly. We conclude by discussing the implications of the role of culture and amenities in neighborhood connectivity for theories of segregation, mobility, and the broader organization of city life, and suggesting pathways for expanding the relational approach developed in this study. More broadly, our results highlight the significant role of culture and amenities in fostering urban cohesion, influencing mobility, social integration, and the spatial organization of city life.

\section*{Results}\label{sec2}

\subsection*{Structure of neighborhood connectivity}

\begin{figure}[h!]
\centering
\includegraphics[width=\linewidth]{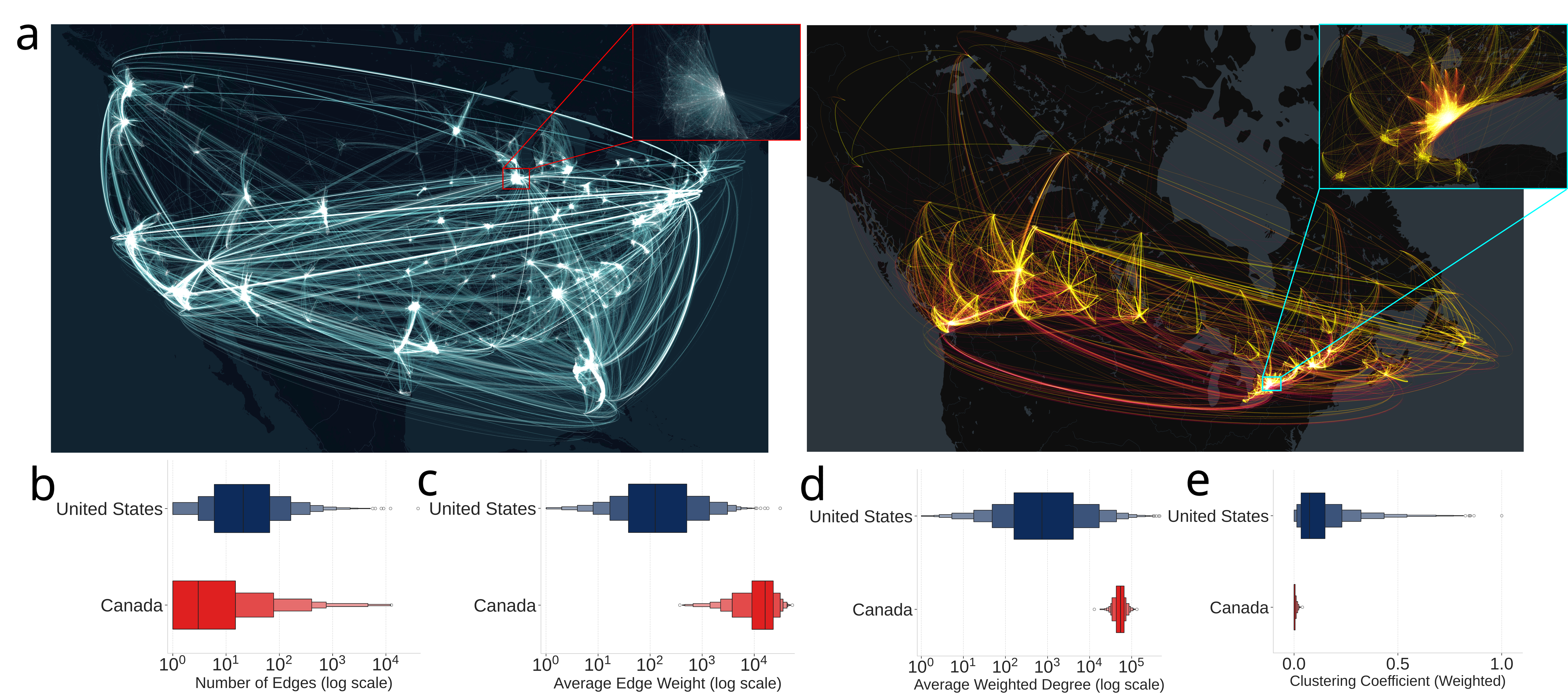}
\caption{Comparative structural properties of urban mobility networks in the United States and Canada. The figure shows the networks studied and the distribution of four network measures across metropolitan areas in each country: (a) visualization of networks studied, the top right figure shows the Canada network (the zoomed part represents the Toronto area), and the top left one shows the U.S. network (the zoomed part represents the Chicago area), (b) number of edges, (c) average weighted degree, (d) average edge weight, and (e) weighted clustering coefficient. The U.S. network, constructed from over 650 million Google Places reviews linking establishments co-visited by the same users, exhibits greater density and higher connectivity, reflecting short-term, voluntary mobility captured through digital traces. In contrast, the Canadian network, derived from nearly 30 million officially registered address changes across three census years, is sparser, with weaker connections and lower clustering, representing long-term residential mobility.}
\label{fig:NetworksAndProperties}
\end{figure}

Using two large longitudinal datasets, we built neighborhood connectivity networks within all metropolitan areas in the USA and Canada (Fig. \ref{fig:NetworksAndProperties}).
The U.S. network captures short-term, consumption-based connections between neighborhood pairs (ZIP codes), constructed from Google Places review data. Co-visitation patterns were derived from users who reviewed at least two establishments. The Canadian network reveals longer-term residential connections between neighborhood pairs (FSAs) and is derived from change-of-address administrative records contained in Canadian tax filings. These networks capture different layers of mobility: the daily circulation of people through places and the more enduring redistribution of residents across urban space.

To uncover structural properties of these networks, four metrics were analyzed for each metropolitan area: the number of edges, the average weighted degree, the average edge weight, and the weighted clustering coefficient (Fig. \ref{fig:NetworksAndProperties} b-e). The observed distributions are consistent with the underlying behavioral mechanisms and modes of data collection. The U.S. network exhibits greater density and higher clustering, reflecting the frequency and different purposes of everyday mobility captured through volunteered digital traces, as well as the overall structure of the U.S. metropolitan system. The Canadian network is (necessarily) sparser and less clustered, yet displays stronger average edge weights where substantial flows of residential relocation occur between major FSAs. These baseline contrasts illustrate how data source and mobility type jointly shape network topology, providing a foundation for interpreting the relational patterns analyzed below. 

\begin{figure}[ht]
\centering
\includegraphics[width=\linewidth]{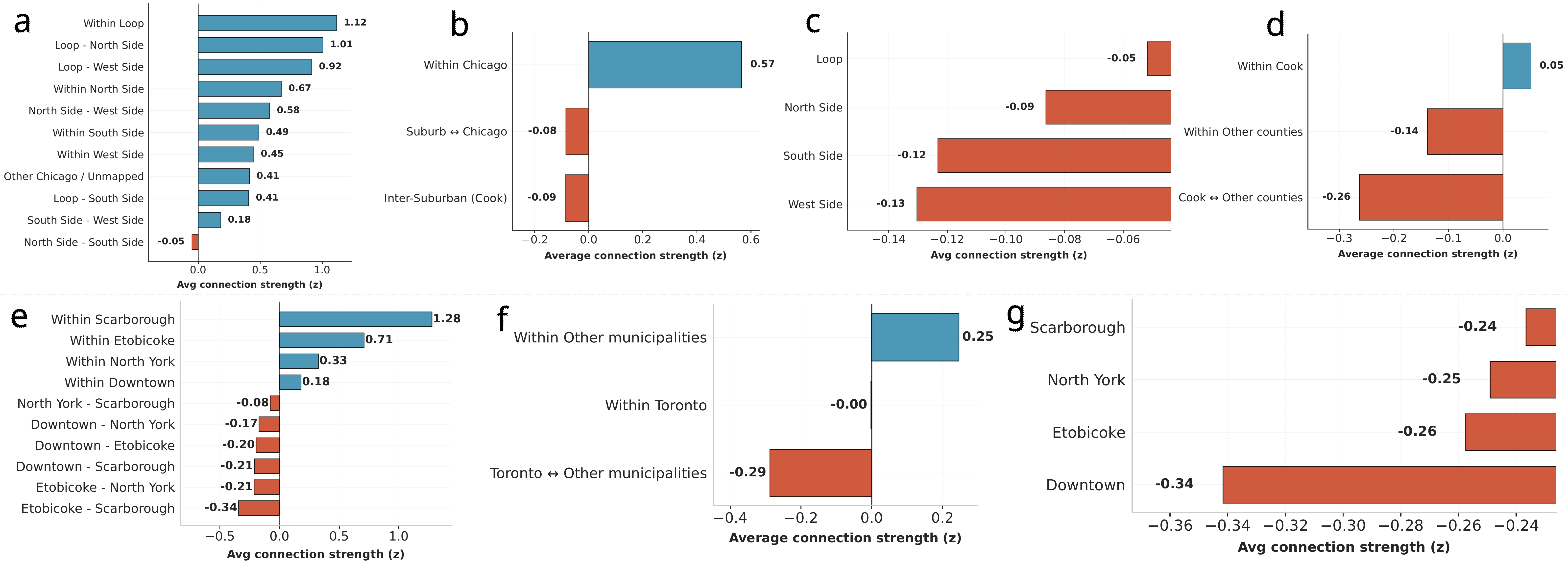}
\caption{Urban connectivity networks in Chicago (top) and Toronto (bottom) reveal contrasting spatial mobility patterns. Each link represents a standardized connection (z-score of edge weight) between neighborhoods based on co-visitation (U.S.) or residential mobility (Canada). Chicago’s network revolves around the Loop and the North-South divide, while Toronto’s shows multiple hubs reflecting a more polycentric structure. }
\label{fig:urbanConnChiTor}
\end{figure}

Connectivity patterns in Chicago and Toronto highlight the plausibility of the data and the kinds of neighborhood networks they reveal, see Fig. \ref{fig:urbanConnChiTor}. In Chicago, the strongest co-visitation connections are within the Loop and between the Loop, the North Side, and the West Side, with additional density within the North Side itself. The relative isolation of the South Side appears in its weaker (though still positive) connectivity to the Loop and in the limited linkage between the North and South sides. At broader scales, visits within the City of Chicago are substantially stronger than those between the city and its suburbs, or among the suburbs themselves. Cook County is markedly more internally connected than its surrounding counties, which remain more self-contained. Overall, Chicago shows a pronounced pattern of centrality: downtown exerts a strong pull, while regional self-containment reinforces spatial hierarchy across scales.

In Toronto, the metropolitan residential network displays a similar structure but reflects a different urban formation and process. Within the city, the enduring strength of its boroughs is clear. Although these municipalities were amalgamated in 1998, they continue to anchor most residential mobility, with moves concentrated within each borough and more rare between them, especially between the far west and east (Etobicoke and Scarborough). Across the wider region, the same pattern persists: moves are more common within municipalities than across them. The city’s inner-suburban boroughs (Scarborough, North York, Etobicoke) are more connected to the surrounding suburbs than is the downtown core. Toronto’s pattern is thus less centered and more polycentric. Residential mobility reveals multiple internal zones of mobility rather than a single dominant hub, yet the structure remains largely consistent across scales.

\subsection*{Relational similarity across multiple domains}

Human mobility does not arise from a single dimension, but from the interplay of multiple relational forces. We begin by examining how similarity across multiple domains covaries across pairs of neighborhoods. This correlational mapping establishes the relational field from which our later multivariate models draw. Classical and recent accounts emphasize the role of geographic, economic, and demographic similarity measures in structuring interaction \cite{schelling1971dynamic,blau1977inequality,wang2018urban,levy2020triple,kraemer2020mapping,candipan2021residence,phillips2021social,huang2023rooted,de2024people,abbiasov202415,xu2025using,jisaGustavo2025,candipan2025neighborhood}, while mobility research has shown that distance strongly constrains movement patterns \cite{gonzalez2008understanding,song2010limits}. Yet cities are also organized around political \cite{doering2021spatial,Huang2023}, cultural, and amenity-based affinities, which have been highlighted in urban studies of consumption and lifestyle clustering~\cite{clark2002amenities}. All of these dimensions coexist and change over time, as does connectivity. 

Most sociological and urban research on neighborhoods in general and on neighborhood connectivity and relationality in particular has focused on demographic and economic variables. This emphasis has largely reflected data availability, since national censuses and household surveys primarily collect information of that kind. As local information about voting patterns has become more widely available,  it is now possible to incorporate political similarity and difference as well, in line with broader discussions about the impact of politics on migration and mobility \cite{Huang2023,gelman2024red}. The emergence of fine-grained Point of Interest (POI) data has added a further dimension, allowing researchers to examine local amenities as indicators of neighborhood activity spaces \cite{silver2016scenescapes,silvaSilver2024EPB}. 

While the amenity mix itself captures the functional composition of local environments, sociologists of culture have emphasized that the meaning of a given amenity may convey additional information about a location’s cultural style or scene, or what the place ``says'' or means. Is this a glamorous or transgressive place, a neighborly or traditionalistic one, a place that prizes ethnic authenticity or corporate brands? Some or all? Different POIs can sustain similar cultural symbolisms, and the same establishment can carry multiple meanings depending on its combinations with others \cite{silver2016scenescapes}. To capture this symbolic level, we measure the cultural ``scene'' of each neighborhood. This measure, detailed in the Supplementary Information (Section \ref{secScenes}), is based on a coding protocol refined in prior research that utilizes systematic human judgment to classify the symbolic meaning of amenities. The considerable predictive power of this measure, which we demonstrate below, serves as a key validation of this approach. 

We extend past work by jointly analyzing both the amenity mix and the inferred cultural meanings of those amenities. This dual approach allows us to separate the functional and symbolic dimensions of local environments and to test how their interaction relates to urban connectivity. 
 
\begin{figure}[ht]
\centering
\includegraphics[width=\linewidth]{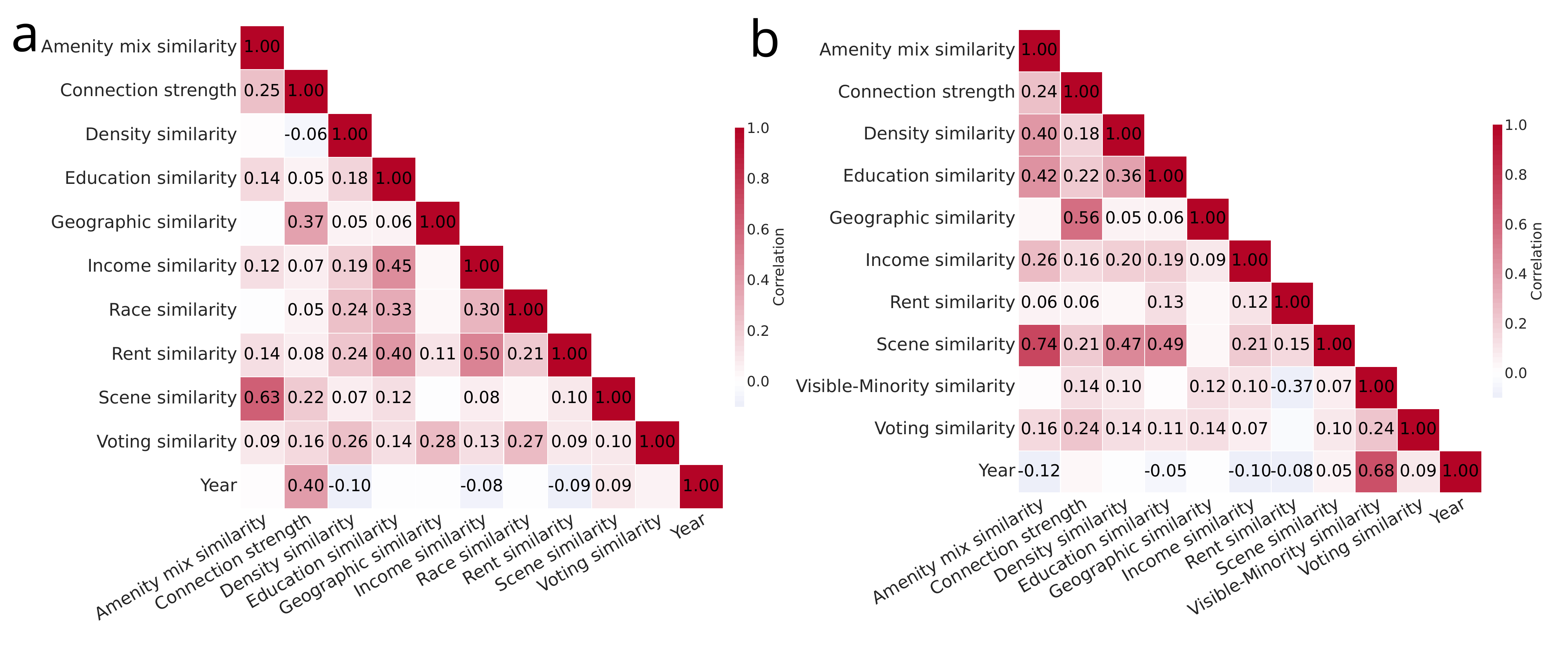}
\caption{Spearman correlation heatmaps show the relationships among demographic, voting, geographic, cultural, and amenity (mix) similarity measures, with \textbf{connection strength} as the focal variable. Results are shown for the United States (a) and Canada (b). Values in the range $[-0.05,+0.05)$ were omitted in the figure.
}
\label{fig:correlations}
\end{figure}

To integrate these perspectives in a way that no previous study has, we compute pairwise similarities across metropolitan neighborhoods in Canada and the United States, spanning demographics, politics, economics, geography, amenities, and culture. The variable connection strength is placed at the center of this analysis. Fig. \ref{fig:correlations} shows Spearman correlation heatmaps among these similarity measures.

In the U.S. dataset, correlations are generally modest, with a few clear patterns. Google Places connection strength is most strongly associated with geographic similarity  ($\rho$ = 0.37). It is also modestly associated with amenity mix similarity ($\rho$  = 0.25), cultural scene similarity ($\rho$  = 0.22), and voting similarity ($\rho$  = 0.16). Demographic and economic similarities are generally positive but are more weakly tied to these co-visitation links. Connection strength tends to grow over time ($\rho$  = 0.4), which may represent platform effects. Our statistical models below correct for this effect.

Other regions of the heatmap reveal how these consumption-based ties fit into a broader structure of urban relationality. Pairs of neighborhoods that are more ethno-racially similar also tend to resemble one another in terms of education, housing costs, income, and voting, forming a race–class–politics alignment (although correlations are all below 0.5). Neighborhoods similar in cultural style share comparable amenity mix, but this cultural–amenities alignment remains partly distinct from the race–class–politics alignment. 
 Geographic similarity matters, but the relation is far from deterministic. Tobler’s “law” operates more as a tendency than as a rule~\cite{sui2004tobler}. 

The Canadian pattern of residential connectivity shows both continuity and contrast. Residential mobility ties are more strongly anchored in geographic similarity ($\rho$  = 0.56), amenities ($\rho$  = 0.24), culture ($\rho$  = 0.21), education ($\rho$  = 0.22), and politics ($\rho$  = 0.24), and exhibit fairly consistent patterns across time.
Neighborhoods with similar cultural styles share similar amenity mix, and these same cultural alignments also coincide with educational, economic, and, to a lesser extent, voting similarity, producing a tighter culture–class alignment. Neighborhoods similar in visible-minority composition exhibit only a weak association with residential connectivity ($\rho$ = 0.14) and a negative correlation with rent similarity ($\rho$ = -0.37); that is, pairs of neighborhoods with similar racial compositions tend to differ in housing costs. This indicates that similarities in racial composition and housing costs are not closely tied. Overall, culture and class are more coupled, while race and class are more distinct, with spatial similarity exerting a somewhat stronger but still not overwhelming impact. 

\begin{figure}[ht]
\centering
\includegraphics[width=\linewidth]{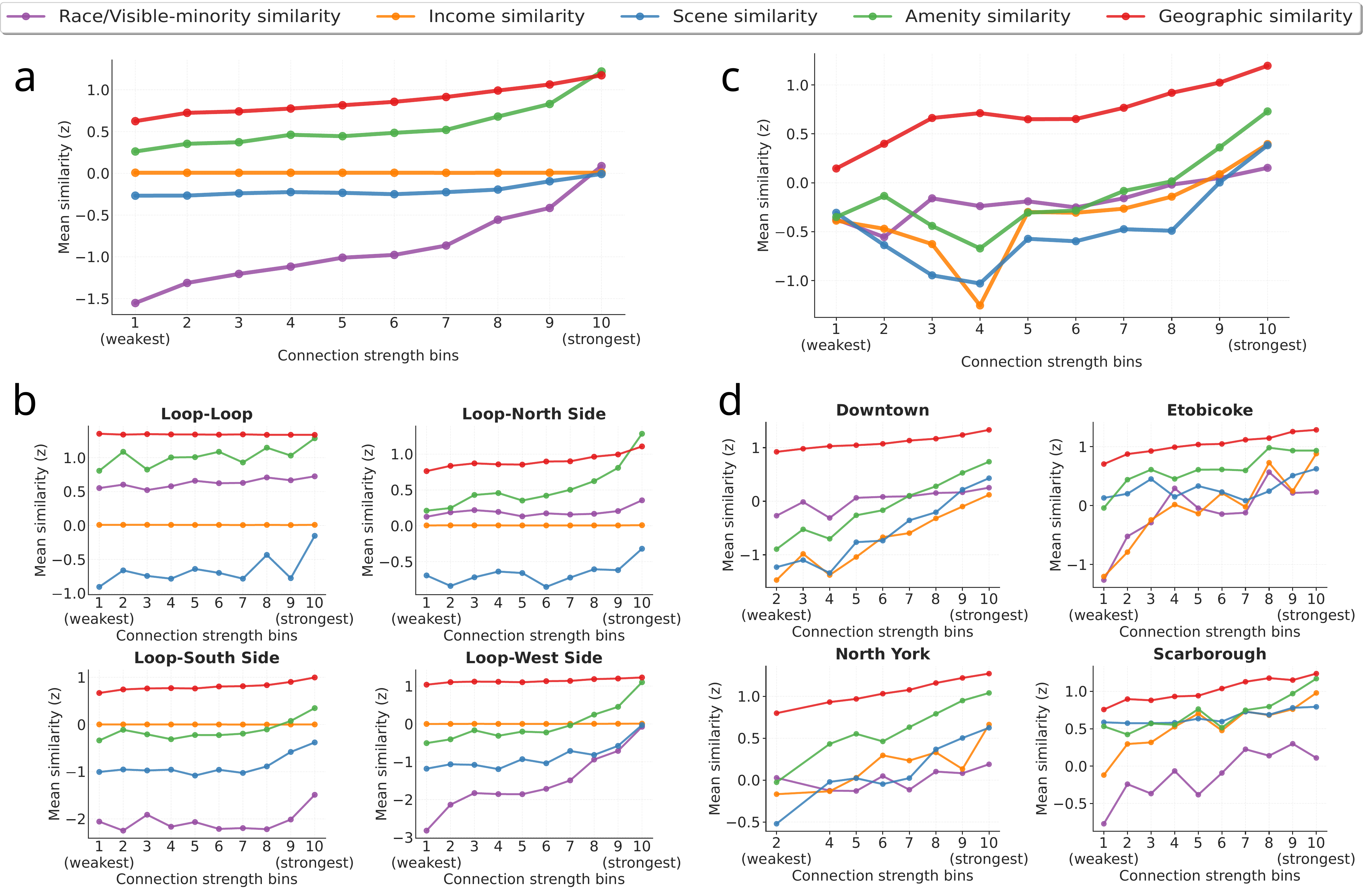}
\caption{Similarity in geography, amenities, culture, income, and race/visible-minority varies systematically across levels of inter-neighborhood connection strength. In Chicago (overall in `a' and specific regions in `b'), geographic similarity and amenity similarity dominate, while cultural and racial homophily increase mainly among the strongest ties. In Toronto (overall in `c' and specific regions in `d'), 
although all similarity measures rise with residential mobility, amenity and culture increase more among the strongest ties.}
\label{fig:examples}
\end{figure}

These national patterns take on clearer form when viewed within specific metropolitan systems. We illustrate this with Chicago and Toronto (Fig. \ref{fig:examples}), which exemplify two contrasting configurations of urban connectivity.

In Chicago (Fig. \ref{fig:examples}-a), overall, at all levels of connectivity, geographic similarity and amenity mix similarity are stronger than income, scene, or racial similarity, though most variables except income increase with tie strength. Amenity mix similarity and race show non-linear curves, rising sharply at the highest levels of connectivity. Racial similarity values hover at or below the national average, underscoring the relative diversity of Chicago compared with other U.S. cities, likely due to the fact that many ties involve trips to and from the Loop, thereby linking ethnically divergent neighborhoods (Fig. \ref{fig:examples}-b). Yet racial dissimilarity is strongest for connections between the Loop and the South or West sides, whereas stronger connections within the Loop and to the North side are more often between racially similar neighborhoods. Taken together, Chicago’s structure reflects both integration and segregation: cultural and amenity homophily increases with mobility, but racial boundaries persist along historical patterns of segregation \cite{sampson2008neighborhood}. 

In Toronto (Fig. \ref{fig:examples}-c), geographic similarity also plays a strong role across all regions and citywide. Amenity and scene similarity rise steeply as neighborhoods become more connected via residential mobility,  particularly within and around the core, while the pattern is more muted in Scarborough (Fig. \ref{fig:examples}-d). Visible-minority similarity also grows with connectivity, but not as steeply, especially in downtown and in North York. Still, the overall picture is one of homophily, where all modes of similarity tend to increase along with growing residential connectivity. 

These patterns provide a descriptive map of the urban relational field. Now, we test these associations more formally to assess how cultural scenes and amenity mix alignments shape connectivity once other factors are controlled.

\subsection*{Multivariate relational models}

To formally test our proposition, we estimate the effects of cultural scenes, amenity mix, and other similarities on neighborhood connectivity using multivariate count models. Because connection strengths are non-negative counts with extremely right-skewed distributions and substantial overdispersion (variance-to-mean ratios of 1,220 in the United States and 81 in Canada), we employ fixed-effects Negative Binomial models, which flexibly accommodate variance far exceeding the mean. As a robustness check, we also estimate Poisson Pseudo–Maximum Likelihood (Poisson) models with the same fixed effects. The results are highly consistent across NB and Poisson estimators (see Supplementary Information, Section \ref{secRobPoisson}).

Our models are designed to answer the question: Which of the addressed variables better explains the number of visits or moves between any two specific neighborhoods? 
We specified node-based dyadic fixed-effects models. This approach is designed to separate the stable, inherent properties of a neighborhood (e.g., baseline popularity and structural position in the city) from the relational factors that connect it to other specific neighborhoods. It achieves this by including a fixed effect for each neighborhood in a pair, which effectively absorbs these time-invariant node-level characteristics. After accounting for the general popularity of the neighborhood pairs, the model can then isolate the effects of the dyadic relationship, such as the pairwise cultural, amenity, or demographic similarities. The models also include year fixed effects to account for national-level trends (such as platform effects or policy changes), and standard errors are clustered by neighborhood pairs to address non-independence. The model's log-link function allows coefficients to be interpreted as (log) proportional changes in the expected count of connections (connection strength). While these are our primary models, supplemental analyses (see Supplementary Information, Sections \ref{secRobYP}, and \ref{secPermut}) confirm the general robustness of the findings presented here.

\begin{figure}[ht]
\centering
\includegraphics[width=\linewidth]{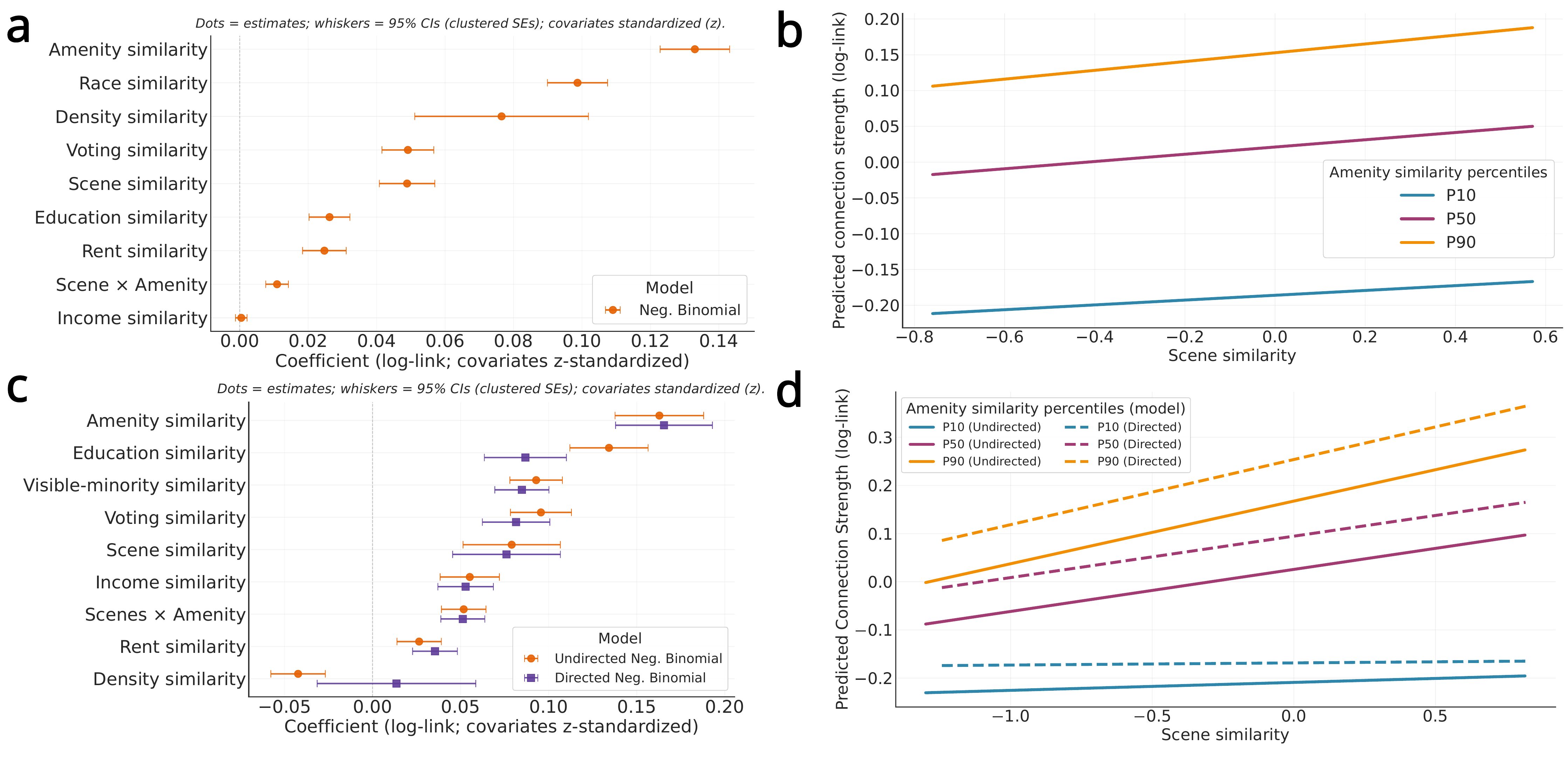}
\caption{Standardized coefficients and marginal effects from multivariate models showing the contributions of spatial,  socioeconomic, voting, amenity, and cultural similarities (as well as their interactions) to inter-neighborhood connection strength. In the figure, we omit the geographic similarity coefficients because they are substantially larger than all other effects: U.S. $b = 0.975$ ($p<0.001$); Canada (undirected) $b = 0.917$ ($p<0.001$); Canada (directed) $b = 0.871$ ($p<0.001$). (a) Estimated standardized coefficients from the U.S. model, based on an undirected Google Places co-visitation network. (b) Log-link of predicted connection strength as a function of scene similarity across low, median, and high levels of amenity similarity in the U.S. (c) Standardized coefficients from Canadian residential-mobility models, comparing undirected and directed (origin–destination) specifications. (d) Log-link of predicted connection strength in Canada as scene similarity varies across amenity-similarity percentiles, shown for both undirected and directed models. Across all countries and model specifications, amenity-mix alignment and cultural similarity emerge as substantial and consistent predictors of inter-neighbourhood connection strength, underscoring their role as core drivers of urban cohesion beyond demographic or spatial proximity.}
\label{fig:coefficients}
\end{figure}

The results for consumption connections (Google co-visitation) in the U.S. are shown in Fig. \ref{fig:coefficients}a. As expected, geographic similarity ($b = 0.975$), omitted in the figure, is the strongest predictor of co-visitation. Supporting our central hypothesis, both amenity similarity ($b = 0.133$) and scene similarity ($b = 0.049$) are positive and significant ($p<0.001$) predictors of connectivity. The model also reveals a significant positive interaction between them ($b = 0.011$). This indicates a reinforcing relationship. As illustrated in Fig. \ref{fig:coefficients}b, the positive effect of shared cultural scenes on co-visitation is stronger when two neighborhoods also share a similar amenity mix. Several demographic and political factors are also statistically significant.
Racial similarity ($b = 0.099$) is a significant predictor, as are similarities in (amenity) density ($b = 0.077$), voting patterns ($b = 0.049$), education ($b = 0.026$), and rent ($b = 0.025$). Income similarity is not significant ($p > 0.60$). 

The model for residential mobility in Canada reveals both parallels and key distinctions (Fig.~\ref{fig:coefficients}c). Geographic similarity ($b = 0.917$), omitted in the figure, remains the dominant factor. The core hypothesis is also supported: amenity similarity ($b = 0.163$) and scene similarity ($b = 0.079$) are both positive and significant predictors of residential ties, and their interaction ($b = 0.052$) is likewise positive and more pronounced than in the U.S. case. At low levels of amenity similarity (10th percentile), shared cultural scenes have virtually no relationship with residential mobility, but the effect becomes strong and positive at medium (50th percentile) and high (90th percentile) levels of amenity alignment (see Fig. \ref{fig:coefficients}d).
This pattern suggests that for relocation decisions, a baseline of functional similarity may be a precondition for cultural resonance. In Canada, income similarity ($b = 0.055$) emerges as a significant predictor of residential mobility,  alongside education ($b = 0.134$), visible-minority composition ($b = 0.093$), voting patterns ($b = 0.096$), and rent ($b = 0.026$). Amenity density similarity has a negative influence ($b =-0.042$). All coefficients are statistically significant. To assess robustness, we also estimate a directed version of the residential mobility network that preserves the origin–destination structure of address changes. The results are substantively similar, confirming that the main findings are not sensitive to the assumption of directionality.

\vspace{1cm}

\section*{Discussion}

The cross-national models confirm that cultural and material factors are significant predictors of neighborhood connectivity for both short-term consumption and long-term residential mobility. These factors operate alongside traditional demographic, economic, and geographic forces. Yet in both national contexts and for both outcomes, amenity mix similarity is the single strongest factor driving neighborhood connectivity, after geographic similarity \cite{gonzalez2008understanding,kraemer2020mapping,schlapfer2021universal,Huang2023,jisaGustavo2025}. Scene similarity is also a substantial factor in both models: cultural similarity is associated with higher connectivity, and is an independent factor in the creation of regular patterns of human mobility.

The relationship between the functions offered by amenities (e.g., the shops, restaurants, services available in an area) and their cultural scene (i.e., whether they express a sense of glamour, tradition, neighbourliness, or ethnic authenticity) is more than additive. In both nations, a positive interaction effect reveals that these two factors amplify one another, though the nature of this amplification differs. For U.S. consumption visits, the effect is one of general reinforcement: shared scenes are always a positive factor, but their attractive power is strongest when two neighborhoods also share similar sets of amenities. For Canadian residential mobility, the interaction suggests a more contingent, two-stage process in which the cultural scene has almost no effect on residential ties when two neighborhoods have dissimilar amenity mixes. Only when a baseline of amenity mix similarity is met does cultural resonance emerge as a significant factor in guiding relocation. This suggests that for high-stakes decisions like moving, people may first sort by practical needs (amenities) and then use cultural style as a finer-grained heuristic to select from within that set. 

The models also clarify the relative role of socioeconomic and political (voting) factors. These distinctions may stem from both the different national contexts (U.S. vs. Canada) and the nature of the outcome variable (short-term consumption vs. long-term residential mobility). In the U.S. model, racial similarity, voting similarity, and rent similarity are all significant predictors of co-visitation, while income similarity is not. This points to a segmentation by race, politics, and housing costs. The Canadian residential model shows somewhat different relative weights across variables. Here, income similarity is a significant factor; similarity in education plays a stronger role, alongside visible-minority composition and voting patterns. However, what stands out is the overall similarity in the patterns of connectivity across the two distinct national and mobility contexts. 

Our study has certain limitations. First, the data-generating processes differ: volunteered digital traces in the U.S. may overrepresent certain users, while Canadian address changes reflect administrative rules and may miss informal moves. Second, measurement error is possible in POI classification and scene inference. Third, the spatial units employed (ZIP codes and FSAs) are relatively coarse, raising modifiable area unit (MAUP) concerns. Although data constraints limit our ability to test alternative neighborhood delineations at finer granularities, results using counties instead of CBSAs, areas considered to study mobilities in the U.S., did not differ substantially (see Supplementary Information, Section \ref{sec:county_robustness_check}).   Fourth, the fixed-effects designs improve internal control but do not establish causality. To broaden coverage and strengthen inference, future work could integrate mobile phone pings, transit data, credit-card transactions, and longitudinal survey microdata, alongside extra qualitative case studies. Combining these with explicit causal designs around exogenous shocks such as transit openings or zoning changes would help identify mechanisms and test the logic of amplification more directly across diverse metropolitan settings.

In conclusion, we began by proposing that urban cohesion also rests on a soft infrastructure of cultural styles or scenes (shared meanings) and amenity mix (material affordances). Our findings validate this specific proposition hypothesis
and underscore the importance of studying local areas within a broader comparative and relational context. We show that social sorting operates powerfully along cultural scenes and amenity mix dimensions, an effect that is amplified when the two align. These forces coexist with, and operate distinctly from, potent sorting along demographic and socioeconomic lines. Amenity and scene similarity may act as heuristics that people use to evaluate unfamiliar areas and guide mobility decisions, allowing them to answer the question: is this a place ``for me'' and am I ``for it.''  

Our analysis provides the outlines of a choice architecture at the metropolitan scale. Mobility (the decision to visit or to move) is a culturally and socially mediated process in which individuals navigate a set of economic factors, symbolic signals, varying mixes of local amenities, and social similarities and differences. Our findings suggest that the weight of these variables differs by context. For discretionary, short-term visits (U.S.), cultural style and amenity mix are reinforcing. For high-stakes, long-term moves (Canada), the process may be more sequential: amenity alignment establishes a ``choice set'' of neighborhoods, and cultural resonance then guides the final selection from within that set. Even so, broadly similar relations shape mobility across divergent contexts and forms. Because these heuristics influence movement, uneven distributions of amenities or scenes can reinforce spatial inequalities and should be considered in urban policy and planning.

More generally, our results support a view of social structure as an emergent outcome of the continual coordination between what places offer, what they mean, who inhabits them, and what they cost. By specifying the distinct and interactive roles of amenities and scenes, we provide a more granular model of how key parts of this coordination happen. The cultural and functional alignment of neighborhoods helps define the boundaries of distinct markets for housing, retail, and services, while also revealing additional aspects of economic acts, such as purchasing a home or frequenting a shop. Expanding this relational approach is crucial for capturing the complex interplay of alignment and differentiation that produces the patterns of connection and division defining modern social life.

\section*{Methods}

\subsection*{Data sources and units of analysis}

We study neighborhood connectivity using two complementary national datasets. For the United States, we construct a panel of short-term mobility based on co-visitation inferred from user review activity, aggregated to ZIP Code Tabulation Areas (or ZIP codes for short). The corpus comprises over 650 million reviews (from 2015 to 2021) that link co-visited establishments, covering all U.S. metropolitan areas. For Canada, we construct a panel of residential moves from confidential Statistics Canada change-of-address records (based on tax filings), aggregated to Forward Sortation Areas (FSAs), spanning three census years (2001, 2006, 2011) and nearly 30 million registered moves, covering all Canadian metropolitan areas. Together, these sources provide complementary perspectives on how neighborhoods are interconnected through both transient circulation through places and longer-term residential relocation at the metropolitan scale.

\subsection*{Network construction}

We represent both systems as weighted networks in which nodes correspond to neighborhoods (ZIP codes in the U.S. and FSAs in Canada) and edges capture the strength of observed connections. For the United States, edges represent the frequency of co-visitation events between neighborhoods based on anonymized Google Places data, yielding an undirected graph of short-term, consumption-based ties. For Canada, we use longitudinal tax-file records that capture address changes between postal areas, constructing an undirected residential mobility network as done for the U.S. case. Because the Canadian data record both origin and destination, we also estimate a directed version of the network as a robustness check. These networks are assembled annually (U.S.) and at each census wave (every five years in Canada), yielding metropolitan panels analyzed separately by country. 

\subsection*{Neighborhood similarity measures}

For each dyad and time, we compute pairwise similarity in:

\begin{itemize}
    \item \textbf{Amenity mix:} cosine similarity of category frequency vectors derived from establishment classifications.
    \item \textbf{Cultural scene:} cosine similarity of scene dimensions that map establishment categories onto theoretically grounded symbolic dimensions (e.g., Corporateness, Charisma, Self-Expression), following prior protocols \cite{silver2016scenescapes,jisaGubert2025,silvaSilver2024EPB}; details of scoring, transfer mapping, and validation are provided in the Supplementary Information, Section \ref{secScenes}.
    \item \textbf{Socioeconomic and politics:} similarity in income, education, rent, visible-minority (Canada) or race composition (U.S.), and voting.
    \item \textbf{Geography and density:} similarity in amenity density and great-circle distance. \end{itemize}

Where applicable, distances/differences are converted to similarities in the range $[0,1]$, to facilitate interpretation across predictors. See Supplementary Information, Section \ref{secSimilarities}, for more details. Pairwise Spearman correlations among these measures are used to produce the correlation heatmaps for each country (Fig.~\ref{fig:correlations}). 

\subsection*{Statistical modeling}

Our dependent variable is the count of connections, \textit{connection strength}, between neighborhood pairs (visits or moves). Because these counts are highly skewed and exhibit substantial overdispersion, we estimate fixed-effects negative binomial regression models with a log link. To isolate relational effects from stable node-level attributes (e.g., population size, centrality, or reputational appeal), we include fixed effects for both neighborhoods and year, and cluster standard errors by dyad to account for dependence within pairs. See Supplementary Information, Section \ref{secModelSpec}, for extra information. As a robustness check, we also estimate Poisson Pseudo–Maximum Likelihood models with the same fixed effects.

In our regression models, we retain the dependent variable in its original scale and standardize only the predictors (mean 0, standard deviation 1). Our main specification standardizes variables at the national level, ensuring the comparability of coefficients across metropolitan areas. We also report models with within-metro-area standardization as a robustness check.

Coefficients indicate proportional changes in expected connections, with an interaction between amenity similarity and scene similarity testing their joint effects. Models are estimated separately for the United States and Canada.

Robustness checks of the model results are presented in the Supplementary Information (See Section \ref{secRobAnalyses}). 

\subsection*{Reporting and replication}

Model estimates are presented as incidence-rate ratios with cluster-robust 95\% confidence intervals. Detailed documentation of preprocessing, similarity measures, and modeling procedures is provided in the Supplementary Information. Upon publication, all code will be released in a public repository, conditional on data-use agreements for the Canadian microdata. 

\subsection*{Data availability}

U.S. establishment and review data are licensed from third-party Google Places review corpora described in \cite{liuctopic2022} and \cite{yan_personalized_2023}, publicly available in \url{https://cseweb.ucsd.edu/~jmcauley/datasets.html#google_local}. We will release ZIP-code–level derived products (annual co-visitation edge weights, amenity frequency vectors, scene vectors, and dyad-level similarity matrices) in a public repository upon publication.

Canadian change-of-address microdata have been accessed through Statistics Canada Research Data Centre (RDC) under an approved project. These confidential microdata cannot be shared; however, we will provide all code and instructions needed to obtain equivalent data from an RDC. 

Socioeconomic covariates for the United States were obtained from IPUMS NHGIS (2017–2021 ACS 5-year estimates; registration required). Canadian 
socioeconomic covariates (2001, 2006, 2011) were obtained from the 2001, 2006, and 2011 Census of Population (Census Profile), publicly released by Statistics Canada. U.S. federal election results (2016, 2020) were sourced from state-released precinct-level returns and are publicly available; we will release the ZIP-aggregated vote shares derived for this study. Canadian federal election results (2000, 2006, 2011) were obtained from publicly available official voting results datasets; we will release the FSA-aggregated vote shares derived for this study.

All geographic boundary files (U.S. TIGER/Line; Canadian CMA/CA, CT, and PCCF-based FSA boundaries) are publicly available from national statistical agencies. Supplementary Information, Section \ref{secDatasetsSup}, provides further details.

\subsection*{Code availability}

All preprocessing, mapping, and modeling code will be made publicly available upon publication in an online repository. The release includes:

\begin{itemize}
    \item parsers for establishment/review data and construction of annual ZIP-level mobility graphs (U.S.);
    \item Research Data Centre (RDC)-compatible scripts for constructing FSA mobility graphs from change-of-address data (Canada), plus public mock pipelines that reproduce all steps on synthetic inputs;
    \item implementations of amenity mix and scene vector constructions, all the similarity computations, and all fixed-effects negative binomial models;  
    \item code to replicate all figures and tables from raw inputs to estimates.
\end{itemize}

Because we cannot distribute Canadian confidential microdata or the raw third-party U.S. review corpus, the repository includes deterministic build scripts that regenerate every derived artifact from (i) public inputs and (ii) RDC-accessible scripts, yielding identical outputs given access to the respective sources. Instructions for obtaining each input and a full environment specification will be provided.

\subsection*{Ethical compliance}
This research utilized publicly available, anonymized Google Places data and restricted-access Statistics Canada microdata, obtained under RDC agreements (all procedures adhered to institutional ethics standards and the confidentiality requirements of Statistics Canada).

\bmhead{Acknowledgments}

This research was supported by funds to the Canadian Research Data Centre Network (CRDCN) from the Social Sciences and Humanities Research Council (SSHRC), the Canadian Institute for Health Research (CIHR), the Canadian Foundation for Innovation (CFI), and Statistics Canada. Although the research and analysis are based on data from Statistics Canada, the opinions expressed do not represent the views of Statistics Canada. This study was also supported in part by the SocialNet project (process 2023/00148-0 of São Paulo Research Foundation - FAPESP), and by the National Council for Scientific and Technological Development - CNPq (processes 314603/2023-9, 441444/2023-7, 313122/2023-7, and 444724/2024-9).
\clearpage


\section*{Supplementary Information}

Supporting the main text, this Supplementary Information provides in Section~\ref{secDatasetsSup} detailed information on geographic divisions, documentation of the data sources (e.g., Google Places reviews, change-of-address records, socioeconomic indicators, election 
votes, and scenes data), as well as the preprocessing procedures. Section~\ref{secNetsCons} describes the construction of the networks; Section~\ref{secSimilarities} outlines the similarity measures and accompanying analyses; Section~\ref{secModelSpec} presents the model specifications. Finally, Section~\ref{secRobAnalyses} reports the robustness analyses.

\section{Datasets}\label{secDatasetsSup}

\subsection{Geographic divisions}

\paragraph{United States: ZIP codes, CBSAs and Counties.}

ZIP Code Tabulation Areas (ZIP codes), used as neighborhood-level units, and Core Based Statistical Areas (CBSAs), areas where mobility is observed, are employed in the main analyses of the U.S. data. Counties are used for robustness checks only.

Geographic boundaries for U.S. analyses were obtained from the 2023 \textit{TIGER/Line Shapefiles}\footnote{\url{https://www.census.gov/geographies/mapping-files/time-series/geo/tiger-line-file.2023.html\#list-tab-790442341}}, which provide official Census Bureau polygons for all administrative divisions. We used shapefiles representing ZIP codes and CBSAs for the metro areas analyzed. We also used shapefiles representing counties for robustness checks (Section~\ref{sec:county_robustness_check}).

The TIGER/Line data already includes associations between ZIP codes and states, but not between ZIP codes and CBSAs, and ZIP codes and counties. To assign ZIP codes to their respective CBSA and county, we used the Missouri Census Data Center’s Geographic Correspondence Engine (Geocorr)\footnote{\url{https://mcdc.missouri.edu/applications/geocorr.html}}. This tool cross-references multiple geographic units and enables reliable neighborhood correspondence.  
Using Geocorr, we successfully matched 33,631 of 33,791 ZIP codes (99.5\%) to CBSAs/counties.

\paragraph{Canada: FSAs, CTs, DAs, CMAs}

Forward Sortation Areas (FSAs), used as neighborhood-level units, and Census Metropolitan Areas (CMAs), areas where mobility is observed, are employed in the main analyses of Canada data. Census Tracts (CTs) and Dissemination Areas (DAs) are used to align the analysis with the available election data. 

Geographic boundaries for Canada were obtained from several official Statistics Canada sources. Polygons for CMAs were derived from the 2011 Census - Boundary files\footnote{\url{https://www12.statcan.gc.ca/census-recensement/2011/geo/bound-limit/bound-limit-2011-eng.cfm}}.   To link postal codes (FSAs) to geographic coordinates and Census areas, we used the Postal Code Conversion File (PCCF), accessed via the University of Toronto’s MDL Library\footnote{\url{https://mdl.library.utoronto.ca/collections/numeric-data/census-canada/postal-code-conversion-file-2011-census-geography}}. The PCCF provides latitude–longitude coordinates and correspondence between postal codes, CTs, DAs, and CMAs.

\subsection{Google Places reviews (United States)}

We use a large-scale Google Places review dataset \cite{liuctopic2022,yan_personalized_2023}, which 
allows capturing co-visitation behavior across U.S. metropolitan areas. The dataset contains 666,324,103 user reviews of 4.96 million establishments written by 113.6 million users nationwide. The data are heavily concentrated between 2015 and 2021, the period we analyze here.
The dataset is organized by state and consists of two main tables: \emph{establishments} and \emph{reviews}.  

\begin{itemize}
    \item \textbf{Establishment information} includes the unique Google Places place identifier, name, address, latitude, longitude, category tags, average rating, number of reviews, price indicator, operating hours, current operational status (open or permanently closed), and a link to its public page.  
    \item \textbf{Review information} includes a unique user identifier, user name, timestamp (Unix epoch time in seconds), rating (1–5), review text, attached images (if any), and any owner responses with corresponding timestamps.  
\end{itemize}

We linked reviews and establishments through their shared Google Maps place identifier and geocoded each establishment to a ZIP Code based on its latitude and longitude. Establishment categories were used to compute each neighborhood’s amenity mix and cultural scene profile (see Section \ref{secScenes}). Review linkages between establishments and users were used to estimate co-visitation networks between ZIP codes, where an edge represents aggregated user overlap (see details in Section \ref{secNetsCons}). Duplicate reviews (8,445,129, or 1.27\% of total) were identified and removed prior to analysis. 
The ones chosen to be removed were not concentrated in a specific state, i.e., eliminations were fairly distributed across states (mean 1.3\%), leaving 657,878,974 unique reviews for subsequent modeling.

\subsection{Change of address data (Canada)}

We use the Canadian Census Health and Environment Cohort (CanCHEC), accessed through the Research Data Centre (RDC), a probabilistically linked dataset that tracks the adult, non-institutional (i.e., household) long-form census population for mortality and cancer outcomes, as well as historical mailing address postal codes. These historical postal codes enable the linkage of environmental data to examine associations between environmental exposures and health outcomes. The linkage was approved by Statistics Canada’s senior management.

The annual postal code history (i.e., mobility file) included in the CanCHECs is constructed primarily from mailing addresses reported in T1 tax files, supplemented with postal codes from other administrative data sources. In practice, the postal code representing a given year reflects the mailing address provided when filing taxes in the following spring (e.g., the postal code recorded for 2000 corresponds to the address submitted with T1 filings in spring 2001). 

After postal codes from tax and administrative sources are compiled, imputation procedures are used to fill gaps in individuals’ postal code histories. For example, as shown in Table \ref{tabExample}, a missing sequence of postal codes from 2002 to 2007 is imputed by assuming stability in residence when the postal codes before and after the gap are identical.

\begin{table}[h!]
\tiny
\caption{Example of imputation of missing postal codes. }
\begin{tabular}{l|l|l|l|l|l|l|l|l}
\hline
                          & \textbf{2001} & \textbf{2002} & \textbf{2003} & \textbf{2004} & \textbf{2005} & \textbf{2006} & \textbf{2007} & \textbf{2008} \\ \hline
\textbf{Raw data}         & K1A0T6       &               &               &               &               &               &               & K1A0T6       \\ \hline
\textbf{After imputation} & K1A0T6       & K1A0T6       & K1A0T6       & K1A0T6       & K1A0T6       & K1A0T6       & K1A0T6       & K1A0T6       \\ \hline
\end{tabular}
\label{tabExample}
\end{table}

We refer to the postal-code mobility file within CanCHEC as the \textit{CanCHEC change-of-address dataset}, which records annual residential moves based on historical mailing-address postal codes.

The CanCHEC change-of-address dataset provides residential mobility information for the 2001, 2006, and 2011 censuses. Each record captures a move between an origin and a destination postal area, which we aggregate to the FSA level. For each census year, we construct a mobility matrix counting all moves occurring within a five-year window (two years before and two years after the census year) across all pairs of FSAs in Canada. These counts form the basis for the inter-neighbourhood mobility networks described in Section~\ref{secNetsCons}. Across the three censuses, the dataset includes 1,253 FSAs and 29,190,445 change-of-address records.

All analyses were conducted in accordance with Statistics Canada's confidentiality, disclosure-control, and data-suppression requirements. The following rules apply:

\begin{itemize}
\item FSAs with fewer than five respondents are excluded.
\item All mobility counts are randomly rounded to the nearest multiple of five, following Statistics Canada conventions.
\item Released counts of zero represent true values between one and five within the same province; we assign these a midpoint value of 2.5, rounded to 3 for use in negative binomial models.
\end{itemize}

Change-of-address records were aggregated into weighted inter-FSA mobility networks for each census year. Edge weights represent the number of moves between neighborhoods.

\subsection{Socioeconomic data}

\paragraph{United States: income, education, rent and race.}

Socioeconomic data for the United States were obtained from American Community Survey (ACS) 5-Year Estimates (Block Groups and Larger Areas), accessed through the IPUMS NHGIS platform \cite{manson_national_2023}. Because they don’t provide yearly data in the ZIP code level, we used the ACS 2013-2017 as a proxy for the year 2015, the ACS 2014-2018 for year 2016, ACS 2015-2019 for year 2017, ACS 2016-2020 for year 2018, ACS 2017-2021 for year 2019, ACS 2018-2022 for year 2020, and ACS 2019-2023 for year 2021. For each year of ACS data, we collected median household income, median gross rent, racial composition (White, Black or African American, Asian, and Hispanic), educational attainment for the population 25 years and over (from which we extract the percentage of bachelor holders), and total population 
 for each ZIP code.

\paragraph{Canada: income, education, rent, and visible-minority.}

Socioeconomic data for Canada were obtained from Statistics Canada’s Census of Population (Census Profile) for 2001, 2006, and 2011, aggregated at the Forward Sortation Area (FSA) level. The following variables were extracted: Housing: Average rent; Demographics: Visible-minority status; Income: Median household income; Population: Total population per FSA; Education: Proportion of population aged 25 to 64 with a university bachelor’s degree or higher. Seven FSAs lacked complete census information and were therefore excluded.

\subsection{Election data}
\label{secSIelection}

\paragraph{United States.}

Precinct-level voting data for the 2016 and 2020 U.S. Presidential Elections were obtained from the MIT Election Data and Science Lab, as documented in \cite{Baltz2022-ek}. These datasets compile and standardize raw precinct returns collected individually from state election authorities or, when necessary, via OpenElections\footnote{https://openelections.net}. For each available precinct, the data include unique geographic identifiers (precinct name, county name, and associated FIPS\footnote{Federal Information Processing Standards} codes), 
candidate names and party affiliations, total votes cast, and the structure of the electoral contest. All data undergo extensive quality-assurance procedures, yielding national presidential vote totals that differ from certified totals by less than 0.01\% in both years.

To aggregate precinct-level results to the ZIP code level, we used an area-weighted interpolation procedure. For each ZIP code polygon, we identified all precincts whose geometries intersected it and apportioned precinct vote totals according to the fraction of each precinct's area falling within the ZIP code. Democratic and total votes were then summed over all intersecting precinct fragments. Finally, the Democratic vote share for each ZIP code was computed as the ratio of aggregated Democratic votes to aggregated total votes.  

Validation checks indicate limited spatial heterogeneity within ZIP codes in both election years, though with small differences between them. In 2016, ZIP codes contained on average 12.6 precincts and precincts intersected 2.39 ZIP codes, with a mean within-ZIP-code standard deviation in Democratic vote share of 0.0888. In 2020, these values were similar—an average of 12.9 precincts per ZIP code and 2.41 ZIP codes per precinct, with a mean within-ZIP-code standard deviation of 0.0868. These low levels of within-ZIP-code variation support the reliability of ZIP code-level estimates as a proxy for local political polarization in both years.

\paragraph{Canada.}

Census Tract (CT)-level voting data for Elections Canada’s federal election results (2000, 2006, 2011) were obtained from publicly available Official Voting Results datasets\footnote{Elections Canada. Official Voting Results (2000, 2006, 2011). Available at \url{www.elections.ca}}. Voting data originally reported at the level of polling districts were spatially harmonized to 2001 Census Tract boundaries. This involved assigning census dissemination blocks to polling districts based on the greatest area overlap and then aggregating block-level assignments up to census tract units. This procedure ensures that election results from different years are comparable within a consistent geographic framework.

To link voting results to FSAs, we used the Postal Code Conversion File (PCCF)\footnote{Available at \url{https://mdl.library.utoronto.ca/collections/numeric-data/census-canada/postal-code-conversion-file}}, which provides correspondence between DAs, CTs, and FSAs. Because CTs may contain multiple DAs that map to different FSAs, votes must be redistributed proportionally. Thus, we used a procedure that consists of these key steps:

\begin{enumerate}[label= Step \arabic*., start=1, leftmargin=1.25cm]

\item \textbf{Link CT-level results to DAs and FSAs.}
Harmonized CT–level federal election results (for 2000, 2006, and 2011) were first matched to 2001 Census Tract identifiers and linked to the PCCF. The PCCF was then used to associate each DA $\mathcal{D}$ with both its parent CT $\mathcal{C}$ and an FSA $\mathcal{F}$. DAs belonging to only one FSA (using the single link indicator (SLI) equal to 1) were selected. Thus, all DAs belonging to the same CT $\mathcal{C}$ inherit the same party-specific voting results  ($V_\mathcal{C}(p)$) from that CT.

 \item \textbf{Define population weights within FSAs and aggregate.}
Party-specific FSA-level vote indicators are obtained as population-weighted aggregates of the inherited CT-level values:
\[
V_\mathcal{F}(p) = \sum_{\mathcal{D} \in \mathcal{F}} \frac{N_\mathcal{D}}{N_\mathcal{F}} \, V_\mathcal{C}(p),
\]
where $p$ is the analyzed party, ${N_\mathcal{D}}/{N_\mathcal{F}}$ is  DA-level population weights within each FSA.  DA-level population counts $N_\mathcal{D}$ are obtained from Statistics Canada's Census\footnote{Publicly available at \url{www150.statcan.gc.ca/n1/en/type/data}}; and 
${N_\mathcal{F}}$ is total population of FSA $\mathcal{F}$, calculated, based on all DAs it has, as
\[
N_\mathcal{F} = \sum_{\mathcal{D} \in \mathcal{F}} N_\mathcal{D}.
\]

 \item \textbf{Compute FSA-level vote shares.}
For each FSA $\mathcal{F}$, the total over all parties is computed as
\[
V_\mathcal{F} = \sum_{p} V_{\mathcal{F}}(p),
\]
FSAs with $V_\mathcal{F} = 0$  are excluded from the analysis to avoid spurious vote shares in areas with negligible or missing electoral data.

The vote share for party $p$ in FSA $\mathcal{F}$ is defined as
\[
S_{\mathcal{F}}(p) = \frac{V_{\mathcal{F}}(p)}{V_\mathcal{F}}.
\]

\end{enumerate}

This procedure is applied separately for 2000, 2006, and 2011, producing FSA-level vote shares for each year.

As a validation of the explained procedure, we evaluated an alternative approach in which each FSA is assigned the vote shares of the geographically closest CT (${\mathcal{C}^*}$) based on great-circle distance between centroids:
\[
S_{\mathcal{F}}(p)  = {\frac{V_{\mathcal{C}^*}(p)}{N_{\mathcal{F}}}.}
\]

Across all years, this alternative method yields Liberal vote-share estimates that are highly consistent with those produced by our population-weighted disaggregation (correlations between 0.93 and 0.95). The absolute differences are modest, with mean deviations ranging from 0.036 in 2001 to 0.026 in 2011 and corresponding standard deviations between 0.047 and 0.034. This confirms that, although the two approaches differ in methodology, the resulting spatial patterns of political support remain closely aligned. We tested both methodologies in our models, and the results were similar. We retain the first approach because it computes the vote counts more precisely.

\subsection{Scenes data}
\label{secScenes}

Scenes data obtainment relies on Scenes Theory. It posits that places convey symbolic meanings, e.g., glamour, transgression, tradition, that attract particular people and activities \cite{silver2016scenescapes}. The Scenes framework operationalizes these meanings along 15 theoretically grounded dimensions grouped under three meta-domains (theatricality, authenticity, legitimacy). For Canada, an additional “Natural” dimension was proposed by \cite{silver2017some}, resulting in 16 total dimensions, all of which are considered in this study. Prior research has shown that scene alignment correlates with residential patterns, political behavior, and economic change \cite{silver2016scenescapes,silvaSilver2024EPB,jisaGubert2025}.

\subsubsection{Measuring cultural scenes}

We distinguish functional composition (amenity categories) from symbolic meaning (scene). Scene dimensions derived from the North American Industry Classification System (NAICS)/business-registry listings and Yellow Pages (YP) data follow the procedures detailed in \cite{SilverClark2016Appendix}. In both cases, amenity categories are mapped to a set of scene dimensions (15 in the U.S.; 16 in Canada) through a coding rubric that assigns each category a score on each dimension. Although the underlying business sources differ in category coverage and amenity counts, the same theoretical framework and scaling conventions are applied to ensure cross-source comparability. In this study, we use the pre‐computed scene scores from both sources (NAICS and YP), following the same documented method to enable comparability across time and data sources.

Following \cite{silver2016scenescapes}, amenity categories are mapped to scene dimensions (15 in the U.S., 16 in Canada) via a multi-step coder protocol involving a tutorial, handbook, and iterative independent coding, yielding high levels of inter-coder consistency. This produces category-level “scene seed” vectors with $D{=}15$ or $D{=}16$  scores per category for NAICS and Yellow Pages datasets. For full details on rubric design, inter-coder reliability, and validation, see \cite{silver2016scenescapes}.

\subsubsection{Neighborhoods’ scenes}

Each establishment $v$ receives a scene vector \(\mathbf{s}^v=\{s^v_1,\dots,s^v_{D}\}\), where $D$ is the number of dimensions, by averaging the seed vectors of its assigned categories. A neighborhood’s scene profile  \(\mathbf{n}=(n_1,...,n_D)\) is the mean across the establishments that belong to it:

\[
\label{eq:cultSig}
n_d = \frac{1}{V}\sum_{v=1}^{V} s^v_d,\quad d=1,\dots,D.
\]

Pairwise cultural similarity between neighborhoods is computed using cosine similarity between their corresponding \(\mathbf{n}\) vectors.

\subsubsection{U.S. scenes vectors: Mapping Google Places categories to scene dimensions}
\label{sec:mapeamentocategoriasscenes}

We reused the cross-platform mapping pipeline introduced in \cite{silvaSilver2024EPB} and \cite{jisaGubert2025} to transfer scene dimension scores from the scenes seed catalog (NAIPS+YP categories) to Google Places (GP) establishments. The procedure involves two semantic transfer stages:

\begin{enumerate}
  \item \textbf{Scene seed $\rightarrow$ Yelp (seed transfer).} 

  Yelp categories are represented as descriptive sentences that capture their full taxonomic context. Because Yelp organizes its taxonomy in four hierarchical levels, each sentence concatenates all categories from the root to the leaf. For example, the lowest-level category \textit{“French Restaurant”} yields the sentence “Food $\rightarrow$ Restaurants $\rightarrow$ French Restaurant.”  Each Yelp sentence is then matched to the most semantically similar category from the original scenes seed catalog. The hierarchical phrasing of Yelp provides richer semantic context for embedding models that support the matching process. Both sets (Yelp sentences and scene seed categories) are embedded using Sentence-BERT \cite{reimers-2019-sentence-bert}, and pairwise cosine similarities are computed. Each Yelp label inherits the scene seed vector of its highest-similarity match (the \textit{argmax} of cosine similarity), following \cite{silvaSilver2024EPB}. This produces a seeded mapping of Yelp categories to multidimensional scene scores representing cultural styles such as “Glamorous,” “Ethnic,” or “Traditionalistic.”

  \item \textbf{Yelp $\rightarrow$ Google Places (cross-platform transfer).} 
  Building on the previous mapping, each Google Places establishment is represented by a descriptive sentence that combines its Google category label(s) with the most specific corresponding Yelp leaf category, i.e., the Yelp categories used in Google Places API requests \citep{jisaGubert2025}. Generic Google Places tags such as \textit{“establishment”} or \textit{“point of interest”} are discarded. Sentence embeddings are computed for all GP and Yelp descriptions using Sentence-Transformers (SBERT), and, for each GP establishment, the most semantically similar Yelp sentence is identified. The corresponding Yelp scene vector is then transferred to that establishment.
\end{enumerate}

Implementation follows SBERT  with the \texttt{all-MiniLM-L6-v2} model, chosen after spot-checks on random samples for semantic coherence. Very short or ambiguous Yelp labels (e.g., single-word cuisine names, such as “German”) formed approximately $5\%$ of cases and were excluded from transfer. The resulting establishment-level scene vectors are averaged within each neighborhood to obtain composite cultural scene profiles (Section \eqref{eq:cultSig}); pairwise cosine similarity between neighborhood scene vectors yields the measure of cultural similarity used in the main models.

Following these procedures,  two open-source tools were developed and released to facilitate reproducibility: a script-based implementation (\url{https://github.com/FerGubert/google_places_enricher}) and a graphical interface version (\url{https://github.com/diegookawa/google_places_enricher_2.0}).

Full theoretical exposition, coding manual, coder rubric, mapping examples, embedding configurations, and validation diagnostics (match quality and sensitivity tests) are available in \cite{silver2016scenescapes}, and transfer pipeline and validation experiments are available in \cite{silvaSilver2024EPB,jisaGubert2025}. We follow the same validated procedures in this study. 
Table \ref{tabExScenes} (adapted from \cite{silver2014cultural}) provides an illustrative example of amenities rated highly for each dimension, as well as how coders were guided in making the ratings.

\begin{table}[htb!]
\tiny
\centering
\caption{Dimensions, Coding Questions, and Associated Amenities. Note. This table shows questions coders asked themselves as they scored each amenity along each of the 16 dimensions of scenes, as well as samples of some of the amenities (out of over 1800) rated highly on each dimension. Note that amenities were weighted both positively and negatively on each dimension, but this table only shows examples of positive weightings.}
\begin{tabular}{p{1.2cm}p{5cm}p{5.5cm}}
\hline
\multicolumn{3}{p{11.7cm}}{\centering\textbf{Legitimacy}}\\ \hline
\multicolumn{3}{p{11.7cm}}{\centering The type of legitimacy promoted by a scene consists in the way it affirms or resists some basis of moral authority, some standard of ethically right or wrong action.}                                                                                                                                                                                                           \\ \hline
\multicolumn{1}{c}{\textbf{Dimension}} & \multicolumn{1}{c}{\textbf{Positive}}                                                                                                             & \multicolumn{1}{c}{\textbf{Examples of highly-rated amenities}}                                                                                                                                        \\
                                       & Do the activities associated with {[}insert name of amenity here{]} assert that the right way to behave is…                                       &                                                                                                                                                                                                        \\
Traditional                            & …according to heritage and/or the models provided by exemplary figures from the past?                                                             & Antique Dealers, Stamps for Collectors, Art Restorations, Calligraphers, Picture Restoring, Bibles, Cemeteries, Crests, Genealogy, Archives, Etiquette                    \\
Utilitarian                            & ...to calculatively extract profit, especially for oneself, pursue disciplined, regular work, efficient consumption, and/or delay gratification? & Money Transfer, Money Order and Services, Productivity Consultants, Internet—Electronic Commerce Solutions Providers, Business Brokers                              \\
Egalitarian                            & …according to norms of universal, reciprocal respect for all persons and peoples?                                                                 & Public Libraries, Women’s Organizations and Services, Social and Human Service Organizations, Social Workers, Social Housing, Courts of Law                                                            \\
Self Expressive                        & …in your own way, originally, uniquely, spontaneously, and/or creatively?                                                                         & Artists—Fine Arts, Decals, Schools—Dramatic Art and Speech, Tattooing, Piercing and Body Art, Fine Arts Schools, Musical Groups                                           \\
Charismatic                            & …determined by the aura around a great leader, religious figure, or star?                                                                         & Advertising—Motion Picture, Designers—Apparel, Haute Couture, Art Galleries, Dealers and Consultants, Fashion Show Producers, Monuments \\ \hline

\multicolumn{3}{p{11.7cm}}{\centering\textbf{Theatricality}}                                                                                                                                                                                                                                                                                                                                                          \\ \hline
\multicolumn{3}{p{11.7cm}}{\centering The type of theatricality promoted by a scene consists in the way it affirms or resists some style of appearance, some way of seeing and being seen.}                                                                                                                                                                                                                            \\ \hline
\multicolumn{1}{c}{\textbf{Dimension}} & \multicolumn{1}{c}{\textbf{Positive}}                                                                                                             & \multicolumn{1}{c}{\textbf{Examples of associated amenities}}                                                                                                                                          \\ \hline
                                       & Do the activities associated with {[}insert name of amenity here{]} assert that the appropriate way to display oneself is …                       &                                                                                                                                                                                                        \\
Glamorous                              & …as shining out, glittering, like gold, fashionably, sparkling?                                                                                   & Schools—Hairdressing and Esthetics, Modeling Agencies, Fashion Show Producers, Pearls, Precious and Semi-Precious Stones, Haute Couture, Jewellers                   \\
Formal                                 & …according to ceremonial and/ritualized, often codified standards of appearance or behavior?                                                       & Clergy, Consulates and Other Foreign Government Representatives, Neckwear, Caps and Gowns, Tuxedos, Etiquette, Ethics and Protocol Lessons                                                             \\
Transgressive                          & ….as offending mainstream culture and values?                                                                                                     & Sex Shops, Entertainment—Adult, Tattooing, Piercing and Body Art, Costumes—Masquerade and Theatrical, Casinos, Nudist Parks, Hemp Stores                                                               \\
Neighbourly                            & …as warmly offering intimate, close, personal connection?                                                                                         & Religious Goods, Libraries, Churches and Other Places of Worship, Religious Organizations, Schools—Academic-Elementary and Secondary                                                                   \\
Exhibitionistic                        & …as a body on display, rather than for onès personality or other individual, non-physical attributes?                                             & Modeling Agencies, Escort Service, Sex Shops, Entertainment—Adult, Nudists Parks, Fashion Show Producers, Night Clubs                                                                                  \\ \hline

\multicolumn{3}{p{11.7cm}}{\centering\textbf{\textbf{Authenticity}}}                                                                                                                                                                                                                                                                                                                                                           \\ \hline
\multicolumn{3}{p{11.7cm}}{\centering The type of authenticity promoted by a scene consists in the way it affirms or resists some way to be real or genuine rather than fake or phony.}                                                                                                                                                                                                                                \\ \hline
\multicolumn{1}{c}{\textbf{Dimension}} & \multicolumn{1}{c}{\textbf{Positive}}                                                                                                             & \multicolumn{1}{c}{\textbf{Examples of associated amenities}}                                                                                                                                          \\ \hline
                                       & Do the activities associated with {[}insert name of amenity here{]} assert that being real rather than fake comes from…                           &                                                                                                                                                                                                        \\
Local                                  & …distinct local roots, a particular place with its own organic customs and practices?                                                             & Farmers Markets, Microbreweries, First Nations Goods, Western Apparel, Flags and Banners, Bed and Breakfasts, Aboriginal Organizations                                                \\
Ethnic                                 & …ethnic roots, unadulterated by foreign traits?                                                                                                  & Tapas, Sushi, Thai Food, Chinese Foods, Mexican Foods, Oriental Goods, First Nations Organizations                                                                                                     \\
Corporate                              & …corporate brands, logos, culture, standardization?                                                                                               & Trade Mark Agents—Registered, Trade Mark Development and Searching, Corporate Image Development Service, New Car Dealers, Advertising Agencies                                                         \\
State                                  & …citizenship, in being a member of a nation and participant in civic life?                                                                        & Flags and Banners, Consulates and Other Foreign Government Representatives, Elected Government Representatives, Embassies                                                                              \\
Rational                               & …cognitive understanding, calculation, rational planning, exercise of the mind?                                                                  & Encyclopedias, Telescopes, Microscopes, Astronomy, Consultants (many types), Engineers (many types), Laboratories, Robotics, Insurance Agents                                                          \\ \hline
Natural                                & ...being natural, arising spontaneously, free from willful planning or social intervention?                                                       & Nudists Parks, Fishing and Hunting, Canoe Trip Outfitters, Nature Centres, Nature Parks, Naturopaths, Yoga, Natural and Organic Foods                                                                  \\ \hline
\end{tabular}
\label{tabExScenes}
\end{table}

\subsection{Cultural similarity beyond amenities}\label{secDistintCultural}

A central contribution of our study is to show that forms of functional compatibility, measured as the relative frequency of amenities, and cultural affinity, which situates those amenities within higher-order cultural dimensions (e.g., Corporateness, Charisma, Self-Expression, Utilitarian), capture distinct signals in mobility explanation.

In Canada, the Spearman correlation of scene similarity (cultural) and amenity similarity considering NAICS data is 0.74 (Pearson's r = 0.68) (details for similarities computation see Section \ref{secSimilarities}). In the U.S., the correlation between these variables is 0.63 (Pearson's r = 0.49). Although these associations are substantial, they are far from collinear, indicating that cultural scenes encode distinct information beyond what can be inferred from amenities alone.

\section{Urban mobility network construction}\label{secNetsCons}

\paragraph{United States.}

To construct the co-visitation network for the United States, we included only users who submitted at least two distinct reviews on the Google Places dataset. This filter yielded 57,848,891 users (50.9\% of all reviewers) with multiple reviews. Within each CBSA, we retained users who reviewed establishments in at least two distinct ZIP codes, ensuring that mobility edges represent within-CBSA co-visitation patterns rather than isolated activity. This second filter yielded 37,784,077 users (33.2\% of all reviewers).

For each CBSA, we constructed a weighted undirected graph $G = (V, E)$, where the set of nodes $V$ corresponds to ZIP codes. An edge $e_{i,j} \in E$ connects ZIP codes $v_i$ and $v_j$ with weight $w_{i,j} \in \mathbb{N}$ indicating the number of unique users who reviewed establishments in both locations. These edges represent shared user mobility across neighborhoods. Self-loops were disregarded in the analyses. 

CBSAs consist of a county or counties with at least one urban core of 10,000 or more inhabitants, plus any adjacent counties with high levels of economic integration.

\paragraph{Canada.}

Mobility networks for Canada were constructed using the CanCHEC change-of-address dataset (which covers the census years 2001, 2006, and 2011). For each census year, we built a weighted undirected graph $G_{C} = (V, E)$ for each CMA, where the node set $V$ represents FSAs, and each edge $e_{i,j} \in E$ connects two FSAs $v_i, v_j \in V$ with weight $w_{i,j} \in \mathbb{N}$ equal to the number of individuals who changed residence between those FSAs, according to CanCHEC change-of-address dataset of a specific census year.

The undirected formulation aggregates mutual flows between FSAs, enabling analyses analogous to those performed on the co-visitation networks constructed for the United States. However, the Canadian data also record both origin and destination postal areas for each move, enabling the construction of a directed network that preserves the directional structure of residential mobility.

Formally, the directed network is defined as $G_{C}^{dir} = (V, E_{dir})$, where each directed edge $e_{i \rightarrow j} \in E_{dir}$ represents the number of moves from origin $v_i$ to destination $v_j$. The corresponding edge weight $w_{i \rightarrow j}$ measures the volume of moves in that direction, enabling models that explicitly incorporate asymmetric residential flows.

To ensure that our results do not depend on the assumption of undirected symmetry, we estimated all relational models using both the undirected and directed versions of the network. Self-loops were disregarded in the analyses. The directed models include origin and destination fixed effects to control, for instance, for overall population size and locational attractiveness of FSAs. In all relational models with Canada data, we restrict the analysis to CMAs that contain at least 10 FSAs, thereby excluding small metropolitan regions from the sample.

\section{Measuring similarities between neighborhoods}\label{secSimilarities}

We compute pairwise similarities between neighborhoods (ZIP codes or FSAs) across several dimensions (geography, education, income, race,  rent, voting, amenity density,  amenity mix,  and culture) to capture multiple aspects of inter-neighborhood resemblance. For all continuous distance or difference measures ($dist$), we convert them into normalized similarity scores using the transformation $1 - dist / \max(dist)$, ensuring unit-free comparability across all metrics. Therefore, all resulting similarity scores lie within [0,1], with no possibility of negative values aside from negligible rounding error.

\subsection{Geographic}

For both the United States and Canada, we calculate the centroid-to-centroid great-circle distance between neighborhoods using internal point centroids from official shapefiles. In the U.S., coordinates are projected from WGS84 to the Contiguous U.S. Albers Equal Area Conic (EPSG:5070), which minimizes distortion across the mainland and islands. Distances are computed in meters using GeoPandas\footnote{\url{https://geopandas.org}}, and then normalized into a geographic similarity score following the above transformation.

\subsection{Education}

For both countries, we compute, for each neighborhood, the absolute difference in the percentage of residents with at least a bachelor’s degree. In the United States, this percentage refers to the population aged 25 and older, whereas in Canada it corresponds to those aged 25 to 64.
\subsection{Income}

For both countries, we compute the absolute difference in average household income between neighborhoods in each year.

\subsection{Race}

For the U.S., the racial similarity between two neighborhoods $A$ and $B$  is calculated following \cite{Huang2023} as:

\[
R_{A,B} = \frac{1}{2}\sum_{i=1}^{n} \left| \frac{N_A(i)}{N_A} - \frac{N_B(i)}{N_B} \right|,
\]

\noindent where  $N_A(i)$ is the population of racial group $i$  (White, Black, Asian, or Hispanic) in neighborhood $A$, and $N_A$ is the total population of $A$. The measure captures the minimal population shift required to equalize racial composition across the two neighborhoods.
For Canada, racial similarity is calculated directly by the difference between percentages of residents identified as visible-minority (non-White, non-Indigenous) in neighborhoods $A$ and $B$, respectively.   

\subsection{Rent}

For the USA, we use median gross rent per neighborhood from the American Community Survey (ACS), and compute the absolute differences between pairs of neighborhoods. 

For Canada, we use the average monthly rent per neighborhood from the Census. Then, we compute the absolute difference in rent values between pairs of neighborhoods.

\subsection{Voting}

We compute the absolute difference between pairs of neighborhoods in the share of votes for the Democratic Party in the U.S. and the Liberal Party in Canada in the nearest election years to the year considered in the U.S. co-visitation or Canada change-of-address datasets.

\subsection{(Amenity) Density}

Amenity density (or density for brevity)  is computed as the number of establishments per square kilometer, derived from neighborhood polygons. For both countries, we compute the absolute difference in densities between two neighborhoods.

\subsection{Amenity (Mix)}

For amenity composition, the cosine similarity between two neighborhoods is calculated based on their normalized frequency vector $\textbf{norm} \left[ \mathbf{f}=(f_{1}, \cdots ,f_{C})\right]$  where  $f_{c}$, $c=1,...,C$,  is the total number of venue category $c$ present in a neighborhood at the time window being considered. 

\paragraph{United States.}
In the United States, the frequency vector $\mathbf{f}$ captures the distribution of Google Places categories for all venues ($C=4569$) present in the dataset. To better capture year-to-year variation, we treat the year of a venue’s first review as its opening year, and the year of its last review as its closing year. Using this proxy, a venue is considered “open” in a given year if that year falls between its first and last review years (inclusive).
\paragraph{Canada.}
In Canadian urban neighborhoods, the frequency vector captures a distribution of business categories derived from two complementary sources: (1) NAICS business registries for 2001 ($C=285$), 2007 ($C=293$), and 2011 ($C=293$), corresponding to the respective census years, and (2) Yellow Pages commercial listings ($C=1472$),  treated as temporally static in robustness checks. 

\subsection{Cultural Scene}

For both countries, we quantify cultural similarity using cosine similarity between neighborhood-level “scene” vectors, following the approach of \cite{silver2016scenescapes}. Each neighborhood is represented as a vector of scene scores derived from its amenity composition, which encodes the neighborhood’s cultural “style”. Scene vectors are computed from the relative frequency of establishment categories in each neighborhood (e.g., music venues, cafés, galleries, gyms) and weighted by category-specific scene typologies as detailed in prior work \cite{silver2016scenescapes,jisaGubert2025,silvaSilver2024EPB}. The resulting vectors express each neighborhood’s alignment along cultural dimensions such as artistic, domestic, or commercial orientations, and their pairwise cosine similarity reflects the closeness of these cultural styles. See Section \ref{secScenes} for additional details on the Scenes Theory framework and data processing steps.

\subsection{Correlation of similarity computations}

Pairwise Spearman correlations among all the previous standardized similarity measures were computed across all inter-neighborhood observations (ZIP–ZIP or FSA–FSA links) in each dataset (U.S and Canada). We also included \textit{year} as a numeric variable; therefore, we capture not only spatial but also temporal co-variation among similarity measures, revealing how these relationships evolve. The resulting correlation matrices, presented in Fig.~\ref{fig:correlations}, summarize the joint spatial and temporal alignment of cultural, socioeconomic, and geographic similarity dimensions. 

To clarify the correlation procedure, the similarity measure for each variable $k$  out of a total $K$ variables \{geographic, education,  ..., amenity, scenes\} is represented as a vector $\mathbf{s}_k$ whose elements correspond to all inter-neighborhood pairs in the dataset
\[
s_k(A,B,t),
\]
where  $s_k(A,B,t)$ is the similarity between neighborhoods $A$ and $B$ for variable $k$ in year $t$.  Pairwise Spearman correlations are then computed between these vectors, such as $\rho(\mathbf{s}_k, \mathbf{s}_j)$, capturing how two variables $k$ and $j$  co-vary across all ZIP--ZIP or FSA--FSA links. The variable \textit{year} is treated analogously as a numeric vector $\mathbf{y}$ with the same dimension as $\mathbf{s}_k$, containing the concatenation of considered years and its pairwise Spearman correlations is calculated as $\rho(\mathbf{s}_k, \mathbf{y})$.

\section{Model specification}\label{secModelSpec}

\subsection{Fixed effects model specification details}\label{secModelEffects}

To estimate the effects of amenity and cultural similarity on neighborhood connectivity while accounting for unobserved, time-invariant heterogeneity, we employed fixed-effects negative binomial models. These models address over-dispersed count data in mobility flows and include neighborhood1, neighborhood2, and year fixed effects. The neighborhood1 and neighborhood2  effects (ZIP code or FSA identifiers) absorb all stable neighborhood characteristics, such as long-term attractiveness, geographic accessibility, or socioeconomic reputation.

For the United States, we estimated the following specification using the \texttt{fenegbin} function from the \texttt{fixest} package in R:

\begin{equation*}
\begin{aligned}
\text{Mobility Strength} \sim{}&
 \text{Income similarity} + \text{Rent similarity} + \text{Race similarity} \\
&+ \text{Geographic similarity} + \text{Density similarity} 
 + \text{Voting similarity} \\
&+ \text{Education similarity} 
 + \text{Amenity similarity} + \text{Scene similarity} \\
 &+ \text{Amenity similarity} \times \text{Scene similarity} \\
&\big|\; \text{neighborhood1} + \text{neighborhood2} + \text{year}.
\end{aligned}
\end{equation*}

For Canada, we used a similar approach, but race similarity was replaced by visible-minority similarity. Additionally, we considered both undirected and directed cases, where, in the directed case, neighborhood1 is the origin and neighborhood2 is the destination. Canadian models were estimated using both NAICS and Yellow Pages data for amenity composition, with the latter serving only as a robustness check (see Section~\ref{secRobYP}).

All models employed cluster-robust standard errors at the dyad level (\texttt{neighborhood1 + neighborhood2}) to account for correlation within neighborhood pairs. 
Except for connection strength (the dependent variable), which remains in its original scale, all predictor variables were z-standardized (mean = 0, SD = 1). Therefore, all predictors were standardized  at the national level to facilitate comparability of coefficients across predictors and national contexts.  Accordingly, the coefficients represent the multiplicative change in the expected count associated with a one-standard-deviation increase in each similarity measure.  For example, a coefficient of 0.20 indicates that a one-SD increase in similarity raises expected flows by $exp(0.20) - 1$ $\simeq$ 22\%. We also report models with within-metro-area standardization as a robustness check (see Section \ref{secCompGloLocal}). 

\subsection{Justification of model choice}\label{secJustification}

The choice of estimator is guided by the distributional properties of the dependent variable, ``connection strength,'' which represents a non-negative count of visits or residential mobility between neighborhoods. We compare three fixed-effects estimators: Linear (OLS), Poisson, and negative binomial (NB).

The linear model is unsuitable as it fails to account for the highly skewed distribution of our count data.  Although the Poisson Pseudo-Maximum Likelihood (Poisson) estimator is consistent even under overdispersion, our data exhibit substantial variance relative to the mean (with variance-to-mean ratios of 1,220 for the United States and 81 for Canada). We therefore adopt the negative binomial estimator as our primary specification, as it explicitly models this overdispersion and offers potential efficiency gains.

This choice is supported by the substantially improved model fit of the negative binomial relative to the Poisson model, indicated by much lower Bayesian Information Criterion (BIC) values. For the United States, the NB BIC is approximately one-third of the Poisson BIC; for Canada, the NB model reduces the BIC by approximately 40\% (see Tables~\ref{tabUSA_NBvsPoisson} and~\ref{tabCanadaNBvsPoissonNAICS}).

\section{Robustness analyses} \label{secRobAnalyses}

\subsection{Robustness check using Poisson model}\label{secRobPoisson}

To assess the sensitivity of the NB results, we estimate Poisson models (Poisson Pseudo–Maximum Likelihood) with high-dimensional fixed effects. As shown on Tables~\ref{tabUSA_NBvsPoisson} and \ref{tabCanadaNBvsPoissonNAICS}, the coefficients for amenity similarity, scene similarity, and their interaction remain positive and statistically significant across all specifications. This consistency confirms that, while the negative binomial provides a superior fit to the data structure, our core findings regarding amenities and scene characteristics are robust to the choice of estimator.

\begin{table}[htb!]
\centering
\caption{Comparison of fixed-effects negative binomial and Poisson models (United States data). Here, n1 and n2 refer to neighborhoods 1 and 2.}
\tiny
\begin{tabular}{p{4cm}cc}
\toprule
 & \textbf{Poisson} & \textbf{Negative binomial} \\
\midrule
\textbf{Dependent variable:} & \multicolumn{2}{c}{\textit{Connection strength}} \\
\midrule
Geographic similarity  & 1.069*** (0.010) & 0.975*** (0.010) \\
Scene similarity  & 0.036*** (0.005) & 0.049*** (0.004) \\
Amenity similarity & 0.090*** (0.005) & 0.133*** (0.005) \\
Income similarity & 0.000 (0.001) & 0.001 (0.001) \\
Rent similarity & 0.022*** (0.003) & 0.025*** (0.003) \\
Education similarity & 0.025*** (0.003) & 0.026*** (0.003) \\
Voting similarity & 0.033*** (0.004) & 0.049*** (0.004) \\
Density similarity & 0.002 (0.006) & 0.077*** (0.013) \\
Race similarity & 0.089*** (0.004) & 0.099*** (0.005) \\
Scene $\times$  Amenity & 0.001 (0.002) & 0.011*** (0.002) \\
\midrule
\textbf{Fixed Effects:} & & \\
n1 & Yes & Yes \\
n2 & Yes & Yes \\
year & Yes & Yes \\
\midrule
Family & Poisson & Neg. Bin. \\
S.E.: Clustered & by n1 \& n2 & by n1 \& n2 \\
Observations & 2{,}713{,}242 & 2{,}713{,}242 \\
Squared Correlation & 0.945 & 0.619 \\
BIC & 77{,}008{,}025 & 25{,}918{,}592 \\
Over-dispersion & -- & 6.535 \\
\bottomrule
\multicolumn{3}{l}{\footnotesize Standard errors in parentheses. All models include year and neighborhoods (ZIP codes) fixed effects.} \\
\multicolumn{3}{l}{\footnotesize Significance levels: *** $p<0.001$, ** $p<0.01$, * $p<0.05$.} \\
\end{tabular}
\label{tabUSA_NBvsPoisson}
\end{table}

\begin{table}[htb!]
\centering
\caption{Comparison of fixed-effects negative binomial and Poisson models (Canada, NAICS data). Here, n1 and n2 refer to neighborhoods 1 and 2.}
\label{tab:canada_naics_models}
\tiny
\begin{tabular}{lcccc}
\toprule
 & \textbf{Undirected NB} & \textbf{Directed NB} & \textbf{Undirected Poisson} & \textbf{Directed Poisson} \\
\midrule
\textbf{Dependent variable:} & \multicolumn{4}{c}{\textit{Connection strength}} \\
\midrule
Income similarity & 0.055*** (0.009) & 0.053*** (0.008) & 0.059*** (0.011) & 0.042*** (0.010) \\
Rent similarity & 0.026*** (0.006) & 0.036*** (0.007) & 0.039*** (0.008) & 0.048*** (0.008) \\
Visible-minority similarity & 0.093*** (0.008) & 0.085*** (0.008) & 0.115*** (0.010) & 0.106*** (0.010) \\
Density similarity & -0.042*** (0.008) & 0.014 (0.023) & -0.037*** (0.009) & -0.013 (0.015) \\
Voting similarity & 0.096*** (0.009) & 0.082*** (0.010) & 0.144*** (0.012) & 0.124*** (0.012) \\
Education similarity & 0.134*** (0.011) & 0.087*** (0.012) & 0.182*** (0.013) & 0.150*** (0.013) \\
Geographic similarity & 0.917*** (0.025) & 0.871*** (0.027) & 1.218*** (0.034) & 1.116*** (0.035) \\
Amenity similarity & 0.163*** (0.013) & 0.166*** (0.014) & 0.208*** (0.017) & 0.193*** (0.017) \\
Scene similarity & 0.079*** (0.014) & 0.076*** (0.016) & 0.048** (0.019) & 0.050** (0.019) \\
Scene $\times$ Amenity & 0.052*** (0.007) & 0.051*** (0.006) & 0.049*** (0.009) & 0.040*** (0.008) \\
\midrule
\textbf{Fixed Effects:} & & & & \\
n1 & Yes & Yes & Yes & Yes \\
n2 & Yes & Yes & Yes & Yes \\
year & Yes & Yes & Yes & Yes \\
\midrule
Family & Neg. Bin. & Neg. Bin. & Poisson & Poisson \\
S.E.: Clustered & by n1 \& n2 & by n1 \& n2 & by n1 \& n2 & by n1 \& n2 \\
Observations & 88{,}374 & 167{,}289 & 88{,}374 & 167{,}289 \\
Squared Correlation & 0.564 & 0.437 & 0.722 & 0.737 \\
BIC & 571{,}443 & 921{,}756 & 948{,}831 & 1{,}303{,}486 \\
Over-dispersion & 3.616 & 3.680 & -- & -- \\
\bottomrule
\multicolumn{5}{l}{\footnotesize Standard errors in parentheses. All models include year and neighborhoods (FSAs) fixed effects.} \\
\multicolumn{5}{l}{\footnotesize Significance levels: *** $p<0.001$, ** $p<0.01$, * $p<0.05$.} \\
\end{tabular}
\label{tabCanadaNBvsPoissonNAICS}
\end{table}

\subsection{Robustness check using Yellow Pages data}\label{secRobYP}

As another robustness check of our findings, we replicated the main models (Canada) using establishment data from the \textit{Yellow Pages} (YP) database, collected in 2009, which provides an alternative yet comprehensive classification of commercial and service establishments in Canada. Unlike the NAICS dataset, whose taxonomy follows a government-defined economic structure, the YP taxonomy reflects how businesses self-identify in the marketplace, often capturing more granular and culturally meaningful categories. As described in \cite{SilverClark2016Appendix}, the cultural mapping procedure for YP categories follows the same rubric used in prior studies, ensuring conceptual and methodological continuity. Using this dataset, we computed using YP categories amenity similarity, scene similarity, and (amenity) density similarity. We then re-estimated the fixed-effects models for both undirected and directed network configurations, mirroring the specifications applied to the NAICS-based analyses. Because data are available for a single year only, we assume that amenity information in 2009 remained constant and apply it to 2001, 2006, and 2011. This constitutes a clear limitation, particularly for further years, given the increasing likelihood of substantive changes in the urban amenity landscape over time.

The results (Table~\ref{tab:fixed_effects_canada}) show a fair degree of consistency between the NAICS- and YP-based models. In both cases, economic and social similarity measures (e.g., income, rent, and education) remain strong positive predictors of connection strength, and the geographic factor continues to dominate, with nearly identical effect sizes across NAICS and YP, confirming the central role of spatial structure in shaping mobility flows.

\begin{table}[!htbp]
\centering
\tiny
\caption{Fixed-effects estimates for Canadian residential mobility networks (NAICS and Yellow Pages data, undirected vs. directed). Here, n1 and n2 refer to neighborhoods 1 and 2. Values in parentheses are standard errors.}
\label{tab:fixed_effects_canada}
\begin{tabular}{lcccc}
\toprule
 & \multicolumn{2}{c}{\textbf{NAICS}} & \multicolumn{2}{c}{\textbf{Yellow Pages (YP)}} \\
\cmidrule(lr){2-3} \cmidrule(lr){4-5}
 & \textbf{Undirected} & \textbf{Directed} & \textbf{Undirected} & \textbf{Directed} \\
\midrule
\textbf{Dependent Var.:} & \multicolumn{4}{c}{\textit{Connection strength}} \\
\midrule
Income similarity & 0.055*** (0.009) & 0.053*** (0.008) & 0.059*** (0.008) & 0.064*** (0.008) \\
Rent similarity & 0.026*** (0.006) & 0.036*** (0.007) & 0.023*** (0.007) & 0.040*** (0.007) \\
Visible-minority similarity & 0.093*** (0.008) & 0.085*** (0.008) & 0.093*** (0.008) & 0.082*** (0.009) \\
Density similarity & -0.042*** (0.008) & 0.014 (0.023) & -0.005 (0.009) & 0.095*** (0.024) \\
Voting similarity & 0.096*** (0.009) & 0.082*** (0.010) & 0.105*** (0.009) & 0.094*** (0.010) \\
Education similarity & 0.134*** (0.011) & 0.087*** (0.012) & 0.193*** (0.011) & 0.156*** (0.011) \\
Geographic similarity & 0.917*** (0.025) & 0.871*** (0.027) & 0.907*** (0.025) & 0.866*** (0.027) \\
Amenity similarity & 0.163*** (0.013) & 0.166*** (0.014) & 0.207*** (0.013) & 0.137*** (0.016) \\
Scene similarity & 0.079*** (0.014) & 0.076*** (0.016) & 0.020 (0.010) & 0.038** (0.013) \\
Scene × Amenity & 0.052*** (0.007) & 0.051*** (0.006) & 0.045*** (0.007) & 0.064*** (0.008) \\
\midrule
Fixed Effects: n1, n2, year & Yes & Yes & Yes & Yes \\
S.E. clustered by n1+n2 & Yes & Yes & Yes & Yes \\
Observations & 88,374 & 167,289 & 88,374 & 167,289 \\
Squared Correlation & 0.564 & 0.437 & 0.563 & 0.423 \\
BIC & 571,443 & 921,756 & 571,096 & 922,750 \\
Over-dispersion & 3.616 & 3.680 & 3.634 & 3.645 \\
\bottomrule
\multicolumn{5}{l}{\footnotesize Standard errors in parentheses. All models include year and neighborhoods (FSAs) fixed effects.}\\
\multicolumn{5}{l}{\footnotesize Significance levels: *** $p<0.001$, ** $p<0.01$, * $p<0.05$.}\\
\end{tabular}
\end{table}

Amenity similarity and its interaction with scene similarity also retain stable, positive, and significant effects across datasets. Scene similarity shows reduced significance in the YP undirected model but remains positive and significant in the directed specification. Although some fluctuations in coefficients appear between the two taxonomies, these differences are anticipated given the assumptions underlying the YP dataset and its more heterogeneous business classification.

These cross-dataset patterns demonstrate that the association between built-environment similarity, cultural similarity, and mobility connectivity is not dependent on NAIC classification system. Replacing the standardized NAICS taxonomy with a self-identified business taxonomy (YP) yields considerable similarities in the results, reinforcing the robustness and generalizability of our framework.

\subsection{Permutation-based robustness test}\label{secPermut}

To assess whether the associations between inter-neighborhood connection strength and similarity measures could arise by chance, we conducted a permutation test in which connection strength values were randomly shuffled within each metropolitan area and year. This procedure preserves the marginal distribution of connection strengths and local network structure while removing any systematic correspondence between movement intensity and pairwise similarity.

We repeated this procedure 100 times and re-estimated the fixed-effects model for each permutation. We then computed, for each variable $k$, the mean ($\mu_{k,\mathrm{perm}}$) and standard deviation ($\sigma_{k,\mathrm{perm}}$) of the resulting coefficients and compared the observed estimates ($\beta_{k,\mathrm{real}}$) to this empirical null distribution: 

\[
z_k = \frac{\beta_{k,\mathrm{real}} - \mu_{k,\mathrm{perm}}}{\sigma_{k,\mathrm{perm}}}. 
\]

Empirical two-tailed $p$-values were derived from the standard normal distribution. Results are summarized in Table~\ref{tabZcomparison}.

The permutation test showed that all major predictors, except for \textit{income similarity} in the United States, deviated considerably from the null distribution, confirming that the observed relationships could not have arisen by chance.

\begin{table*}[htb!]
\centering
\tiny
\caption{Permutation-based robustness results for the USA and Canada models. 
Empirical $z$-scores and two-tailed $p$-values compare real model coefficients 
(from fixed-effects negative binomial estimation) to the null distribution generated 
from 100 within–metro-year permutations. }
\label{tab:perm_results}
\begin{tabular}{lrrrrr}
\toprule
\textbf{Country} & \textbf{Variable} & \textbf{$\beta_{k,\mathrm{real}}$} & \textbf{$\mu_{k,\mathrm{perm}}$} & \textbf{$\sigma_{k,\mathrm{perm}}$} & \textbf{$z$-score (empirical $p$)} \\
\midrule
\multirow{10}{*}{\textbf{USA}}
 & Geographic similarity & 0.9751 & 0.00003 & 0.00229 &
   425.21 ($p<10^{-15}$) \\
 & Scene similarity & 0.0489 & 0.01969 & 0.00202 &
   14.49 ($p<10^{-15}$) \\
 & Amenity similarity & 0.1330 & 0.00853 & 0.00287 &
   43.30 ($p<10^{-15}$) \\
 & Income similarity & 0.00045 & -0.00135 & 0.00090 &
   2.00 ($p=0.045$) \\
 & Rent similarity & 0.0247 & 0.00735 & 0.00182 &
   9.55 ($p<10^{-15}$) \\
 & Education similarity & 0.0262 & -0.00807 & 0.00182 &
   18.81 ($p<10^{-15}$) \\
 & Voting similarity & 0.0491 & -0.00782 & 0.00193 &
   29.47 ($p<10^{-15}$) \\
 & Density similarity & 0.0765 & 0.03445 & 0.00543 &
   7.75 ($p=9.2\times10^{-15}$) \\
 & Race similarity & 0.0987 & 0.00258 & 0.00210 &
   45.78 ($p<10^{-15}$) \\
 & Scene $\times$ Amenity & 0.0109 & -0.00031 & 0.00049 &
   22.65 ($p<10^{-15}$) \\

\midrule

\multirow{10}{*}{\textbf{Canada}}
 & Income similarity & 0.0552 & 0.00175 & 0.00846 & 
   6.32 ($p=2.6\times10^{-10}$) \\
 & Rent similarity & 0.0264 & 0.00025 & 0.00913 & 
   2.87 ($p=4.1\times10^{-3}$) \\
 & Visible-minority similarity & 0.0929 & -0.00429 & 0.00827 & 
   11.75 ($p<10^{-15}$) \\
 & Density similarity & -0.0423 & -0.00004 & 0.01077 & 
   -3.92 ($p=8.8\times10^{-5}$) \\
 & Voting similarity & 0.0956 & 0.00293 & 0.00913 & 
   10.15 ($p=3.2\times10^{-24}$) \\
 & Education similarity & 0.1342 & 0.00011 & 0.01030 & 
   13.02 ($p<10^{-15}$) \\
 & Geographic similarity & 0.9170 & 0.00085 & 0.01277 & 
   71.72 ($p<10^{-15}$) \\
 & Amenity similarity & 0.1628 & 0.00035 & 0.01324 & 
   12.27 ($p<10^{-15}$) \\
 & Scene similarity & 0.0790 & -0.00426 & 0.01633 & 
   5.10 ($p=3.4\times10^{-7}$) \\
 & Scene $\times$ Amenity & 0.0517 & -0.00167 & 0.00674 & 
   7.92 ($p=2.3\times10^{-15}$) \\

\bottomrule
\end{tabular}
\label{tabZcomparison}
\end{table*}

\subsection{Comparing global vs.\ CMA/CBSA standardization}\label{secCompGloLocal}

To assess the robustness of our findings to different scaling choices, instead of using a global standardization, we re-estimated all models using predictors standardized within each CBSA in the United States and within each CMA in Canada, rather than using global (national) standardization.

Notice that under the global standardization used so far, each predictor's input is transformed using a single mean and standard deviation computed from all observations across all CMAs/CSBAs and all years. A global z-score, therefore, reflects how extreme a value is in the context of the entire dataset, enabling direct comparisons between cities.

Under the within-CMA/CBSA standardization, each predictor is standardized separately for each CMA or CBSA. For example, for Toronto, we compute the mean and standard deviation using only Toronto’s observations, and then scale all Toronto values using those Toronto-specific parameters. The same procedure is applied independently to Vancouver, Montreal, New York, Los Angeles, and every other metropolitan area. A within-area z-score, therefore, captures how unusual a value is relative to that area’s own distribution, abstracting away differences in overall scale between cities. Table~\ref{tab:fixed_effects_usa_global_cbsa} presents the results for the U.S. ZIP-code network, while Tables \ref{tableGlobalCMA_NAIC} and \ref{tableGlobalCMAYellow} report the corresponding estimates for NAICS- and YP-based similarity measures in Canada. 

\begin{table}[h!]
\centering
\tiny
\caption{Fixed-effects estimates for U.S. co-visitation networks, comparing global vs. CBSA-level standardization. Values in parentheses are standard errors.}
\label{tab:fixed_effects_usa_global_cbsa}
\begin{tabular}{lcc}
\toprule
 & \textbf{Global Standardization} & \textbf{CBSA Standardization} \\
\midrule
\textbf{Dependent Var.:} & \multicolumn{2}{c}{\textit{Connection strength}} \\
\midrule

Geographic similarity
    & 0.975*** (0.010) & 0.910*** (0.008) \\

Scene similarity
    & 0.049*** (0.004) & 0.030*** (0.004) \\

Amenity similarity
    & 0.133*** (0.005) & 0.147*** (0.005) \\

Income similarity
    & 0.001 (0.001) & 0.004 (0.003) \\

Rent similarity
    & 0.025*** (0.003) & 0.010*** (0.003) \\

 Education similarity
    & 0.026*** (0.003) & 0.023*** (0.003) \\

Voting similarity
    & 0.049*** (0.004) & 0.052*** (0.004) \\

Density similarity
    & 0.077*** (0.013) & -0.004 (0.005) \\

Race similarity
    & 0.099*** (0.005) & 0.093*** (0.004) \\

Scene × Amenity
    & 0.011*** (0.002) & 0.005** (0.002) \\
\midrule

Fixed Effects: n1, n2, year & Yes & Yes \\
S.E. clustered by n+n2 & Yes & Yes \\
Observations & 2{,}713{,}242 & 2{,}712{,}619 \\
Squared Correlation & 0.619 & 0.876 \\
BIC & 25{,}918{,}592 & 25{,}953{,}112 \\
Over-dispersion & 6.535 & 6.392 \\
\bottomrule
\multicolumn{3}{l}{\footnotesize Standard errors in parentheses. All models include year and neighborhood (ZIP code) fixed effects.} \\
\multicolumn{3}{l}{\footnotesize Significance levels: *** $p<0.001$, ** $p<0.01$, * $p<0.05$.} \\
\end{tabular}
\end{table}

\begin{table}[h!]
\centering
\tiny
\caption{Fixed-effects estimates for Canadian residential mobility networks using NAICS-based similarity measures, comparing global vs. CMA-level standardizations. Values in parentheses are standard errors.}
\label{tableGlobalCMA_NAIC}
\begin{tabular}{lcccc}
\toprule
 & \multicolumn{2}{c}{\textbf{Global Standardization}} & \multicolumn{2}{c}{\textbf{CMA Standardization}} \\
\cmidrule(lr){2-3} \cmidrule(lr){4-5}
 & \textbf{Undirected} & \textbf{Directed} & \textbf{Undirected} & \textbf{Directed} \\
\midrule
\textbf{Dependent Var.:} & \multicolumn{4}{c}{\textit{Connection strength}} \\
\midrule

Income similarity
    & 0.055*** (0.009) & 0.053*** (0.008)
    & 0.052*** (0.008) & 0.053*** (0.008) \\

Rent similarity
    & 0.026*** (0.006) & 0.036*** (0.007)
    & 0.027*** (0.006) & 0.034*** (0.006) \\

Visible-minority similarity
    & 0.093*** (0.008) & 0.085*** (0.008)
    & 0.099*** (0.008) & 0.084*** (0.007) \\

Density similarity
    & -0.042*** (0.008) & 0.014 (0.023)
    & -0.034** (0.012) & 0.041* (0.017) \\

Voting similarity
    & 0.096*** (0.009) & 0.082*** (0.010)
    & 0.075*** (0.008) & 0.062*** (0.007) \\

Education similarity
    & 0.134*** (0.011) & 0.087*** (0.012)
    & 0.130*** (0.011) & 0.081*** (0.011) \\

Geographic similarity
    & 0.917*** (0.025) & 0.871*** (0.027)
    & 0.839*** (0.020) & 0.836*** (0.021) \\

Amenity similarity
    & 0.163*** (0.013) & 0.166*** (0.014)
    & 0.164*** (0.013) & 0.152*** (0.014) \\

Scene similarity
    & 0.079*** (0.014) & 0.076*** (0.016)
    & 0.061*** (0.014) & 0.070*** (0.015) \\

Scene × Amenity
    & 0.052*** (0.007) & 0.051*** (0.006)
    & 0.047*** (0.006) & 0.047*** (0.006) \\

\midrule
Fixed Effects: n1, n2, year & Yes & Yes & Yes & Yes \\
S.E. clustered by n1+n2 & Yes & Yes & Yes & Yes \\
Observations & 88,374 & 167,289 & 88,374 & 167,279 \\
Squared Correlation & 0.564 & 0.437 & 0.573 & 0.557 \\
BIC & 571,443 & 921,756 & 571,589 & 916,832 \\
Over-dispersion & 3.616 & 3.680 & 3.589 & 3.822 \\
\bottomrule
\multicolumn{5}{l}{\footnotesize Standard errors in parentheses. All models include year and neighborhood (FSA) fixed effects.} \\
\multicolumn{5}{l}{\footnotesize Significance levels: *** $p<0.001$, ** $p<0.01$, * $p<0.05$, . $p<0.1$.} \\
\end{tabular}
\end{table}

\begin{table}[h!]
\centering
\tiny
\caption{Fixed-effects estimates for Canadian residential mobility networks using Yellow Pages (YP) data, comparing global vs. CMA-level standardizations. Values in parentheses are standard errors.}
\label{tableGlobalCMAYellow}
\begin{tabular}{lcccc}
\toprule
 & \multicolumn{2}{c}{\textbf{Global Standardization}} & \multicolumn{2}{c}{\textbf{CMA Standardization}} \\
\cmidrule(lr){2-3} \cmidrule(lr){4-5}
 & \textbf{Undirected} & \textbf{Directed} & \textbf{Undirected} & \textbf{Directed} \\
\midrule
\textbf{Dependent Var.:} & \multicolumn{4}{c}{\textit{Connection strength}} \\
\midrule

Income similarity
    & 0.059*** (0.008) & 0.064*** (0.008)
    & 0.057*** (0.008) & 0.061*** (0.008) \\

Rent similarity
    & 0.023*** (0.007) & 0.040*** (0.007)
    & 0.022*** (0.006) & 0.037*** (0.007) \\

Visible-minority similarity
    & 0.093*** (0.008) & 0.082*** (0.009)
    & 0.099*** (0.008) & 0.080*** (0.008) \\

Density similarity
    & -0.005 (0.009) & 0.095*** (0.024)
    & -0.003 (0.010) & 0.079*** (0.020) \\

Voting similarity
    & 0.105*** (0.009) & 0.094*** (0.010)
    & 0.084*** (0.008) & 0.071*** (0.008) \\

Education similarity
    & 0.193*** (0.011) & 0.156*** (0.011)
    & 0.186*** (0.011) & 0.147*** (0.010) \\

Geographic similarity
    & 0.907*** (0.025) & 0.866*** (0.027)
    & 0.827*** (0.020) & 0.831*** (0.021) \\

Amenity similarity
    & 0.207*** (0.013) & 0.137*** (0.016)
    & 0.187*** (0.012) & 0.147*** (0.014) \\

Scene similarity
    & 0.020 (0.010) & 0.038** (0.013)
    & 0.025* (0.010) & 0.032** (0.012) \\

Scene × Amenity
    & 0.045*** (0.007) & 0.064*** (0.008)
    & 0.045*** (0.006) & 0.061*** (0.006) \\

\midrule
Fixed Effects: n1, n2, year & Yes & Yes & Yes & Yes \\
S.E. clustered by n1+n2 & Yes & Yes & Yes & Yes \\
Observations & 88,374 & 167,289 & 88,374 & 167,279 \\
Squared Correlation & 0.563 & 0.423 & 0.574 & 0.559 \\
BIC & 571,096 & 922,750 & 571,278 & 917,093 \\
Over-dispersion & 3.634 & 3.646 & 3.604 & 3.813 \\
\bottomrule
\multicolumn{5}{l}{\footnotesize Standard errors in parentheses. All models include year and neighborhood (FSA) fixed effects.} \\
\multicolumn{5}{l}{\footnotesize Significance levels: *** $p<0.001$, ** $p<0.01$, * $p<0.05$.} \\
\end{tabular}
\end{table}

Overall, the results are fairly consistent across the two standardization strategies in both countries. For instance, geographic similarity remains the strongest predictor of mobility and interest ties in all specifications and scenes, amenity mix, and their interaction also remain positive and statistically significant under both global and within-metro standardization.

\subsection{CBSA vs. county robustness checks}
\label{sec:county_robustness_check}

The U.S. results are also robust to the choice of geographic aggregation used to construct similarity measures. Table~\ref{tab:fixed_effects_usa_cbsa_county} compares models based on CBSA-level versus county-level distributions, both using globally standardized predictors. Across specifications, the signs and significance levels of all key predictors remain highly consistent, including scene similarity, amenity mix similarity, race composition, and education similarity. Some  coefficients (e.g., rent and voting) vary slightly in magnitude across the two definitions, but the overall pattern remains similar.

\begin{table}[h!]
\centering
\tiny
\caption{Fixed-effects estimates for U.S. co-visitation networks under global standardization, contrasting CBSA- and county-based similarity measures. Values in parentheses are standard errors.}
\label{tab:fixed_effects_usa_cbsa_county}
\begin{tabular}{lcc}
\toprule
 & \textbf{CBSA-based Model} & \textbf{County-based Model} \\
\midrule
\textbf{Dependent Var.:} & \multicolumn{2}{c}{\textit{Connection strength}} \\
\midrule

Geographic similarity
    & 0.975*** (0.010) & 0.843*** (0.012) \\

Scene similarity
    & 0.049*** (0.004) & 0.054*** (0.005) \\

Amenity similarity
    & 0.133*** (0.005) & 0.133*** (0.005) \\

Income similarity
    & 0.001 (0.001) & 0.000 (0.001) \\

Rent similarity
    & 0.025*** (0.003) & 0.007* (0.004) \\

Education similarity
    & 0.026*** (0.003) & 0.026*** (0.004) \\

Voting similarity
    & 0.049*** (0.004) & 0.029*** (0.004) \\

Density similarity
    & 0.077*** (0.013) & 0.061*** (0.008) \\

Race similarity
    & 0.099*** (0.005) & 0.092*** (0.005) \\

Scene × Amenity 
    & 0.011*** (0.002) & 0.010*** (0.002) \\

\midrule
Fixed Effects: n1, n2, year & Yes & Yes \\
S.E. clustered by n1+n2 & Yes & Yes \\
Observations & 2{,}713{,}242 & 1{,}091{,}251 \\
Squared Correlation & 0.619 & 0.038 \\
BIC & 25{,}918{,}592 & 11{,}416{,}165 \\
Over-dispersion & 6.535 & 9.548 \\
\bottomrule
\multicolumn{3}{l}{\footnotesize Standard errors in parentheses. All models include year and neighborhoods (ZIP code) fixed effects.} \\
\multicolumn{3}{l}{\footnotesize Significance levels: *** $p<0.001$, ** $p<0.01$, * $p<0.05$.} \\
\end{tabular}
\label{tab:fixed_effects_usa_cbsa_county}
\end{table}

\clearpage


\end{document}